\documentclass[twocolumn]{aastex631}

\shortauthors{Williams et al.}

\usepackage{amsmath}
\usepackage[caption=false]{subfig}
\usepackage{mwe}
\usepackage{multirow}
\usepackage{graphicx}
\begin{document}

\title{$\Lambda$CDM star clusters at cosmic dawn: stellar densities, environment, and equilibrium}

\correspondingauthor{Claire E. Williams}

\author[0000-0003-2369-2911]{Claire E. Williams}
\affil{Department of Physics and Astronomy, UCLA, Los Angeles, CA 90095}
\affil{Mani L. Bhaumik Institute for Theoretical Physics, Department of Physics and Astronomy, UCLA, Los Angeles, CA 90095, USA\\}
\email{clairewilliams@astro.ucla.edu}

\author[0000-0002-9802-9279]{Smadar Naoz}
\affil{Department of Physics and Astronomy, UCLA, Los Angeles, CA 90095}
\affil{Mani L. Bhaumik Institute for Theoretical Physics, Department of Physics and Astronomy, UCLA, Los Angeles, CA 90095, USA\\}

\author[0000-0002-4227-7919]{William Lake}
\affil{Department of Physics and Astronomy, UCLA, Los Angeles, CA 90095}
\affil{Mani L. Bhaumik Institute for Theoretical Physics, Department of Physics and Astronomy, UCLA, Los Angeles, CA 90095, USA\\}

\author[0000-0001-5817-5944]{Blakesley Burkhart}
\affiliation{Department of Physics and Astronomy, Rutgers, The State University of New Jersey, 136 Frelinghuysen Rd, Piscataway, NJ 08854, USA \\}
\affiliation{Center for Computational Astrophysics, Flatiron Institute, 162 Fifth Avenue, New York, NY 10010, USA \\}

\author[0000-0003-3816-7028]{Federico Marinacci}
\affiliation{Department of Physics \& Astronomy ``Augusto Righi", University of Bologna, via Gobetti 93/2, 40129 Bologna, Italy\\}
\affiliation{INAF, Astrophysics and Space Science Observatory Bologna, Via P. Gobetti 93/3, I-40129 Bologna, Italy\\}

\author[0000-0001-8593-7692]{Mark Vogelsberger}
\affil{Department of Physics and Kavli Institute for Astrophysics and Space Research, Massachusetts Institute of Technology, Cambridge, MA 02139, USA\\}

\author[0000-0001-7925-238X]{Naoki Yoshida}
\affiliation{Department of Physics, The University of Tokyo, 7-3-1 Hongo, Bunkyo, Tokyo 113-0033, Japan}
\affiliation{Kavli Institute for the Physics and Mathematics of the Universe (WPI), UT Institute for Advanced Study, The University of Tokyo, Kashiwa, Chiba 277-8583, Japan}
\affiliation{Research Center for the Early Universe, School of Science, The University of Tokyo, 7-3-1 Hongo, Bunkyo, Tokyo 113-0033, Japan}

\author[0000-0002-0311-2206]{Shyam H. Menon}
\affiliation{Department of Physics and Astronomy, Rutgers, The State University of New Jersey, 136 Frelinghuysen Rd, Piscataway, NJ 08854, USA \\}
\affiliation{Center for Computational Astrophysics, Flatiron Institute, 162 Fifth Avenue, New York, NY 10010, USA \\}

\author[0000-0002-8859-7790]{Avi Chen}
\affiliation{Department of Physics and Astronomy, Rutgers, The State University of New Jersey, 136 Frelinghuysen Rd, Piscataway, NJ 08854, USA \\}

\author[0000-0002-8192-8091]{Angela Adamo}
\affiliation{Astronomy Department, Stockholm University \& Oskar Klein Centre, Roslagstullsbacken 21, Stockholm, SE-10691, Sweden.\\}



\begin{abstract}
The {\it James Webb Space Telescope (JWST)} has opened a window on many new puzzles in the early Universe, including a population of high-redshift star clusters with extremely high stellar surface density, 
suggesting unique star formation conditions
in the Universe's early evolution. 
We study the formation and evolution of these first star clusters and galaxies using 
an {\tt AREPO} cosmological simulation box designed to resolve the intricate environments of the smallest halos hosting Population III star clusters at $z \geq 12$.
Our approach, which prioritizes baryonic structure identification through a friends-of-friends algorithm,
 provides new insights into early star cluster formation and delivers predictions directly relevant to observations. 
We investigate the dynamical properties of these first star clusters and use numerical and analytical methods to understand
the populations of virialized and non-virialized systems. 
Our findings indicate that high-$z$ star clusters in a feedback-free regime can achieve extreme surface densities, consistent with the systems detected by {\it JWST}.
These results imply that {\it JWST } may have the opportunity to uncover stellar systems at high redshift whose dynamical state preserves evidence of the hierarchical structure formation process. 

\end{abstract}


\section{Introduction} \label{sec:intro}
The earliest phase of galaxy formation is driven by the hierarchical assembly of dark matter halos and the formation of star clusters from the primordial gas of the early Universe \citep[e.g.,][]{vogelsberger_cosmological_2020}.
The {\it James Webb Space Telescope (JWST)} is revealing both the products of this process--early galaxies at extremely high redshift \citep[e.g.,][]{Finkelstein+23ceersI, Castellano+22glass, Adams+23,adams_discovery_2023,franco_unveiling_2024}--and in a few lensed scenarios, the building blocks themselves \citep[e.g.,][]{Mowla+24,Adamo+24, Vanzella+23}. 
These observations probe the first starlight to light up our Universe at ``Cosmic Dawn," and offer an opportunity to test models of cosmology, star formation, and galaxy buildup.

In the local, low-redshift context, the conversion of molecular gas clouds into stars is well-understood to be an inefficient process, limited by various feedback processes that prevent total collapse \citep[e.g.,][]{mckee_theory_2007}. 
However, at high-redshift, {\it JWST} observations reveal stellar systems of extremely high density. 
For example, {\it JWST } observations find a striking overabundance of compact, blue galaxies 
\cite[e.g.,][]{Finkelstein+23ceersI, Atek+23candidates, finkelstein_complete_2024, cullen_ultraviolet_2024,harikane_jwst_2024}, with resolved sizes implying surface densities consistent with compact elliptical galaxies, ultracompact dwarfs, and superstar clusters in the local context \cite[e.g.,][]{Casey+24, baggen_small_2024}.
Below the scale of galaxies, gravitational lensing effects provide a view of individual star clusters at extremely high redshifts: in particular the ``Firefly Sparkle" at $z=8.3$ \citep{Mowla+24} and the ``Cosmic Gems" at $z=10.1$ \citep[][]{Adamo+24}. 
Additionally, observations at intervening redshifts reveal small clusters \cite[i.e.,][]{ Vanzella+23, Vanzella+22,fujimoto_glimpse_2025,whitaker_discovery_2025}, 
as well as clumps within lensed galaxies \citep[e.g.,][]{mestric_exploring_2022, claeyssens_star_2023} providing an evolving picture of star clusters at various cosmological epochs, including  before, during, and after Reionization.
When individual high-redshift stellar clusters are observed at the parsec scale, the stellar component is seen to be arranged in extremely clumpy, high-density configurations with surface densities exceeding $10^4-10^5 M_\odot\text{ pc}^{-2}$ \citep[e.g.,][]{Adamo+24}. 
Thus, it seems that the conditions for star formation in the first billion years must have differed from the local paradigm.

In the theoretical picture, there are many unique conditions at Cosmic Dawn that differentiate this epoch from standard star formation. 
For one, the first generation of stars or ``Population III" (Pop III) forms from pristine gas, devoid of all metals \citep[e.g.,][]{Yoshida+06}. 
The nature of these first stars, particularly their initial mass function (IMF), is not yet constrained, with some models and simulations predicting high mass cutoffs upwards of $10^3 M_\odot$ \citep[e.g.,][]{Chon+22}. 
Consequently, the impact of feedback processes that may result from such an extreme population is also unknown. 
However, high surface densities observed in the {\it JWST} data thus far already hint at a different mode of star formation during this period, one which allows for highly dense clusters to form.

On the scales of individual star clusters, high-resolution star formation simulations can resolve individual stars and their formation environment given a set of physical assumptions \citep[e.g.,][]{Lake+24b, Chon+22}. 
Zoom-in simulations can achieve excellent resolution and study the clumps of star formation, although they typically do not provide a large, statistical sample of systems \citep[e.g.,][]{kimm_formation_2016, barrow_first_2017, ma_simulating_2018,katz_probing_2019,arata_starbursting_2020,gelli_stellar_2020, kohandel_velocity_2020,2024ApJ...967L..28M}. 
This study focuses on cosmological hydrodynamical simulations, which resolve the broader environment of dark matter halos and galaxies. 
Many large-scale cosmological simulations study the properties of high redshift galaxies \citep[e.g.,][]{shen_thesan_2024, pillepich_simulating_2018,roper_first_2022, marshall_impact_2022,wu_photometric_2020,vogelsberger_cosmological_2020,vogelsberger_high-redshift_2020,vogelsberger_introducing_2014,vogelsberger_properties_2014}; however their resolution is usually on the order of hundreds of pc, making it difficult to study individual star clusters. 
A handful of studies have both the resolution and the statistical sample size to study the population and environment of clusters \citep[e.g.,][]{nakazato_merger-driven_2024,Mayer+24}. 
Here, we investigate a high-resolution cosmological simulation box with thousands of dark matter halos above a minimum resolved halo mass of $M_{\rm DM}=2.3\times 10^5 h^{-1}M_\odot$, which resolves baryonic structure above $M_{\rm b}\sim 2\times 10^4M_\odot$. 

Even the higher resolution simulation suites usually rely on subgrid models, tuned to observations and/or higher resolution simulations, to incorporate the conversion of gas into stars and the desired feedback effects \citep[e.g.,][]{Marinacci+19, smith_grackle_2017}. 
Additionally, in order to make relevant comparisons to observables, it is necessary to group the simulated particles into structures corresponding to galaxies and clusters. 
Many halo- and subhalo-finding algorithms exist in the literature \citep[see e.g.,][]{Knebe+11, Knebe+13}{}{}. 
The simulations studied in this work were run using the {\tt AREPO} simulation code \citep{weinberger_arepo_2020}, and thus, we employ the built-in friends-of-friends (FOF) algorithm, which is described in detail in \cite{springel_populating_2001}. 
We test the method of primarily searching for stars and gas--mimicking an observer that has no knowledge of the underlying dark matter field--to identify early star clusters and protogalaxies. 
This method is most similar to the clump finding method of \cite{nakazato_merger-driven_2024}, although we do not differentiate between young and old stars for the purpose of structure detection.

Here, we investigate the properties of star clusters and early low-mass galaxies in a simulation suite in {\tt AREPO} using friends-of-friends algorithms focused on the dark matter halos and additionally using the baryonic particles as the primary particle type to mimic an observation. 
Since our cosmological simulation does not incorporate the stream velocity, stellar structures should not form outside of dark matter halos in the nominal picture of structure formation.
We find that the two methods lead to discrepant number counts and cluster properties above the resolution scale. We identify the environments leading to this discrepancy in simulations.
Using these methods, we investigate the dynamical state and environment of these primordial galaxies. 
We find and characterize several populations of star clusters by their virialization and suggest that in {\it JWST}'s earliest observing epochs, some systems still preserve the dynamical evidence of the first structure formation.  
Our results imply that the stellar components of the high-z structure are frequently not virialized, and this assumption should not be used to estimate properties in this epoch. 
Additionally, our star clusters are consistent with extremely high surface density, as observed in the Cosmic Gems, Sunrise, and Sunburst clusters at high redshift. 
Because no feedback was included in this simulation box, this implies that 
these clusters may have formed in a scenario where feedback is weak with respect to gravity \citep[e.g.,][]{dekel_efficient_2023}, as explored in recent studies that show high star formation efficiency in high surface density cases \citep[e.g.,][]{menon_outflows_2023,grudic_when_2018,kim_modeling_2018}

The paper is organized as follows. 
In \S~\ref{sec:methods}, we describe the simulation runs, FOF algorithms, and the methods used to test for virialization and boundedness. 
In \S~\ref{sec:boundclusters} we present our findings, comparing the number counts using the different FOF runs. 
We identify two populations of virialized clusters in \S~\ref{sec:virializedclusters}, and use analytical timescale arguments to understand their formation. 
With these numerical and analytical tools in hand, we investigate the overall environment and properties of star clusters in \S~\ref{sec:earlyclusterenv}, comparing to {\it JWST } observations.
A discussion and summary of the work are presented in \S~\ref{sec:summary}.
In this study we assume a $\Lambda$CDM cosmology, with $\Omega_{\rm \Lambda} = 0.73$, $\Omega_{\rm m} = 0.27$, $\Omega_{\rm b} = 0.044$, $\sigma_8  = 1.7$, and $h = 0.71$.

\section{Simulations and identification of star clusters} \label{sec:methods}

\subsection{Cosmological Simulation} \label{sec:box}
For this study, we use the {\tt AREPO} cosmological simulation code to run a ($2.5$ Mpc $h^{-1}$)$^{3}$ volume simulation with $768^3$ particles from $z=200$ to $z=12$. 
Our simulation does not include any feedback effects nor the primordial baryon-dark matter streaming velocity. 
We note that the simulation is the same as the non-stream velocity ($0\sigma_{\rm bc}$) run from \citet[][]{Lake+23b} and \citet[][]{Williams+24}. 

In {\tt AREPO} hydrodynamics is discretized with a  Voronoi mesh that is allowed to move with the fluid flow \citep[][]{Springel2010a}. 
Our minimum simulated Voronoi cell volume is $0.645$ pc$^3$, giving a minimum gas cell size of $\sim0.86$ pc.  
The simulation gravitational softening length results in a minimum particle resolution of $1$~pc at the time of the last snapshot. 

Star formation is implemented according to the method of \cite{Marinacci+19}.
When gas mesh cells exceed the Jeans mass on the scale of the cell, they are eligible to form stars. 
Additionally, star formation is restricted to only gravitationally bound regions--so the virial parameter of the cell must be negative. 
Once these collapse criteria are met, the creation of star particles occurs stochastically, by converting a gas cell into a collisionless star particle with a probability determined by the local star formation rate.
The mass of each star particle is equal to the mass of its progenitor gas cell.
The collapse timescale is set to the gravitational dynamical time of the gas cell ($t_{\rm dyn} = \sqrt{3\pi/(32G \rho_{\rm g})}$).
The local star formation rate of each gas cell is computed as $\dot{M}=\epsilon M_{\rm gas}/t_{\rm dyn}$, where $M_{\rm gas}$ is the mass of the gas mesh cell and $\epsilon$ is an efficiency factor set to $0.01$.

\subsection{Baryonic-centered structure finder} \label{sec:FOF_methods}

\begin{figure*}
    \centering
    \includegraphics[width=\linewidth]{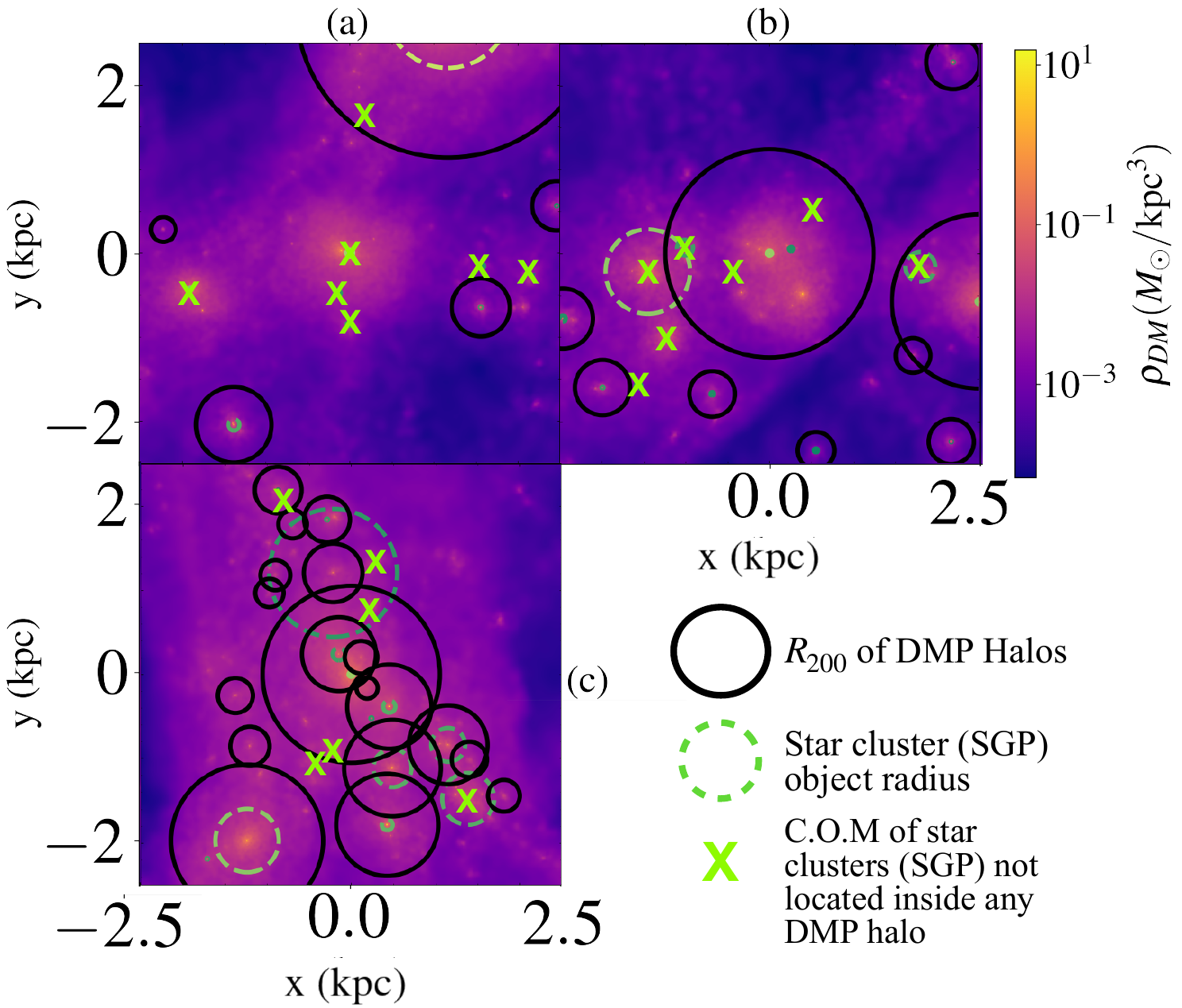}
    \caption{Example systems where DM primary fails to characterize the stellar component. The projected density of dark matter is shown in a 10 kpc region of the simulations. Black circles show $R_{200}$ of dark matter halos as detected by the Dark Matter Primary (DMP) FOF algorithm.
    The green circles show Stars \& Gas Primary (SGP) FOF objects, colored by stellar mass. 
    For non-virialized objects, the FOF radius is shown as a dashed line. 
    For objects that are virialized, the virial radius is shown as a solid line. An ``X" marker is plotted where a star cluster is outside of any DM halo.
    Three cases are shown. {\bf (a)} Several baryonic clusters exist in a region where the underlying dark matter distribution does not meet the DM FOF criteria. 
    {\bf (b)} A star cluster is accreting onto a larger DM halo, and its host is not picked up as a separate halo.   {\bf (c)}  A highly dense environment at the intersection of gaseous and dark matter filaments contains many halos and star clusters. Some star clusters lie outside of the halo $R_{200}$.  
    }
    \label{fig:visuals}
\end{figure*}

 In {\tt AREPO}'s FOF algorithm \citep{springel_populating_2001},  simulation particles are grouped together using a linking-length criterion, where the distance between particles in a prospective group is compared with a multiple of the mean inter-particle separation. 
Traditionally, dark matter is used as the primary linking particle type to identify halos that contain galaxies. 
Each halo is associated with its gravitational radius, often chosen to be the radius at which the halo density falls below $200$ times the mean or critical density at that redshift.  
In the case of {\tt AREPO}'s algorithm, a secondary particle FOF algorithm can then be run to identify the group's other particle types, such as gas and/or stars, providing the stellar mass, gas mass, and star formation rate of galaxies in the simulation box \citep[][]{weinberger_arepo_2020}. 
This method aligns with the typical picture of $\Lambda$CDM galaxies--a small baryonic component of gas and stars sitting at the center of a large, spherical, and cuspy dark matter halo \citep[e.g.,][]{BarkanaLoeb+01}. 
This assumption can reliably track the concentrations of stellar light that would be ``observed" by a telescope under the condition that all significant clusters of stars reside in resolved, approximately spherical dark matter halos. 

Of course, electromagnetic observatories such as {\it JWST} cannot directly study the underlying dark matter field and in this work, we study a context where this assumption breaks down: the earliest epoch of star cluster formation in a high-resolution cosmological simulation. 
This environment is challenging for a dark-matter-focused FOF method because these first structures form in low-mass minihalos, which are subject to frequent mergers, fragmentation, and accretion onto larger structures.
Star clusters may reside in small halos, which are not the subhalos of a larger host halo in the box. Although most cosmological simulations target the dark matter for their catalogs, a method searching for baryonic particles is necessary for many simulation contexts. 
For example, \cite{nakazato_merger-driven_2024} use a ``clump finder" to identify stellar clumps in mock images of the FirstLight simulation. 
In zoom-in simulations, satellite galaxies are sometimes found using search criteria targeted at maxima in the stellar density field \citep[e.g.,][]{gelli_stellar_2020}.
Also, previous works investigating the cosmic streaming velocity searched for gas particles independently of dark matter in order to find structures that were removed from their parent halos by the cosmic steam velocity \citep[e.g.,][]{Popa+15, Chiou+18,Chiou+19,Chiou+21,Lake+21,Lake+22,Lake+23b,Williams+23,Williams+24}. 

We ran a set of FOF structure finding algorithms targeting different particle types on the simulation snapshot outputs at $z=12,13,14,$ and $15$ with {\tt AREPO}'s default halo finding algorithm,  which uses a Friends-of-Friends (FOF) technique \citep[see][for details]{Springel+05}. 
Table~\ref{tab:FOFruns} lists all the runs.
With this method, the user chooses a simulated particle type (typically DM) to serve as the ``primary",  and the FOF outputs a catalog  
of halos. 
The user can optionally select additional particle types (typically stars and/or gas) to serve as the ``secondary," and a secondary stage of the algorithm will associate these particles with the existing halos.
In this work, we ran the standard FOF algorithm that uses dark matter particles as the primary particle type and baryons as the secondary, in addition to three sets of runs with combinations of particles as the primary particle type: DM$+$stars$+$gas, stars$+$gas, and stars only. 
Figure~\ref{fig:visuals} shows the dark matter density in three example regions where the structures identified by these algorithms are overlaid.
The $R_{200}$ of halos detected using the DM-primary, gas$+$stars secondary method are shown in black, while the radii of objects detected by the stars$+$gas method are shown in green. 
Green ``X" markers denote the center of mass of baryon-centric objects which do not lie within the $R_{200}$ of a DM-detected halo. 
From this region, we can see that these methods--which trace concentrated structures in different particle types--will give different catalogs of halos/galaxies.  

As we discuss further in \S~\ref{sec:boundclusters}, there is a similar behavior among algorithms that use DM as one of the primary types, as well as among those that do not. However, important differences exist between these two categories. 
Thus, we refer to ``DM-centric" and ``baryon-centric" structure finding algorithms. 
As denoted in Table~\ref{tab:FOFruns}, when a particular run is not specified, we use the DM primary, stars$+$gas secondary as the default ``DM-centric" dataset, and the stars$+$gas run as the default ``baryon-centric" dataset (because it includes the added information of the gas particles). 

For DM-centric runs, we enforce a resolution cutoff of 300 DM particles. For baryon-centric runs, we enforce a resolution cutoff of 100 star particles. 
We test our results to ensure convergence of the structure-finding algorithm by varying the linking length. 
The results of that investigation are presented in Appendix~\ref{app:convergence}.
A pressure floor is not needed for {\tt AREPO } 
because the gas cells use a Voronoi mesh method. 
In order to assure that our smallest structures are not artifical fragments, we have checked the simulation box to ensure that the Jeans mass is resolved by at least 10 cells.
\begin{table*}[]
\centering
        \begin{tabular}{|l|l|l|l|}
            \hline
            \ttfamily {\bf Category} & \ttfamily {\bf Description} & \ttfamily {\bf Primary Particle(s)} & \ttfamily {\bf Secondary Particle(s)}\\                
            \hline 
            \hline
            \multirow{3}{*}{\bf DM-centric} & \multicolumn{1}{l}{DM primary $^\dagger$ (standard)} & \multicolumn{1}{|l}{Dark Matter} & \multicolumn{1}{|l|}{Stars \& Gas} \\\cline{2-4}
                                & \multicolumn{1}{l}{Dark Matter \& baryons} & \multicolumn{1}{|l}{Dark Matter, Stars,} & \multicolumn{1}{|l|}{None} \\
                                & \multicolumn{1}{l}{primary} & \multicolumn{1}{|l}{and Gas} & \multicolumn{1}{|l|}{ } \\\cline{2-4}
                                \hline
                \multirow{2}{*}{\bf Baryon-centric}& \multicolumn{1}{l}{Stars \& gas primary$^\dagger$} &\multicolumn{1}{|l}{Stars and Gas} & \multicolumn{1}{|l|}{None} \\\cline{2-4}
                                 & \multicolumn{1}{l}{Star primary} &\multicolumn{1}{|l}{Stars} & \multicolumn{1}{|l|}{None} \\\hline
        \end{tabular}
    \caption{Summary of structure finding algorithms used in this work. $^\dagger$ indicates the algorithm used as the standard for the category in the analysis after \S~\ref{sec:boundclusters}, where only the difference between the DM-centric and the baryon-centric methods are considered. In other words, when  we refer to ``DM-centric" and ``baryon centric" algorithms in those sections, we mean the DM primary and the stars\& gas primary runs, respectively. } 
    \label{tab:FOFruns}
\end{table*}

\subsection{Boundedness and Virialization}
\label{sec:boundmethods}

For baryon-centric structure finding runs without dark matter, we check to ensure boundedness to exclude potential numerical artifacts. 
First, among the $N$ star particles in the group, which each have mass $m_{i}$, position $\vec{r_i}=(x_i, y_i, z_i)$, and velocity $v_i = |\vec{v}_i|$ 
we calculate the energy of the star particles:
\begin{equation}
    E_* = \sum_i^N \frac{m_i v_i^2}{2}-\sum_{i}^N\sum_{j\neq i}^N \frac{G m_i m_j}{|\vec{r_i}-\vec{r_j}|} \ .
\end{equation}

If $E_*<0,$ the star particles are certainly bound, and we include that structure in our further analysis.
For groups that fail this criterion, we perform an additional calculation 
incorporating  
nearby dark matter particles to see if the stars are bound to an underlying dark matter halo. 
First, we take all dark matter particles within a sphere surrounding the center of mass of the object with a radius equal to the galactocentric distance of the most distant star particle. 
Among this collection of $N$ stars and $M$ dark matter particles, we calculate the total energy among the stars and dark matter:
\begin{multline}
    E_{*,DM} = E_* + \sum_i^M \frac{m_{dm,i} v_{dm,i}^2}{2} 
    \\ 
    - \sum^M_i\sum_{j\neq i}^M \frac{G m_{dm,i} m_{dm,j}}{|\vec{r}_{dm,i}-\vec{r}_{dm,j}|}
    \\ - \sum_{i}^M \sum_{j}^N \frac{G m_{dm,i} m_{*,j}}{|\vec{r}_{dm,i}-\vec{r}_{*,j}|}.
\end{multline}
If this quantity is negative, we consider the structure to be bound. 
Structures failing this additional criterion are excluded from the analysis as unresolved numerical artifacts. 

\begin{figure}
    \centering
    \includegraphics[width=.99\linewidth]{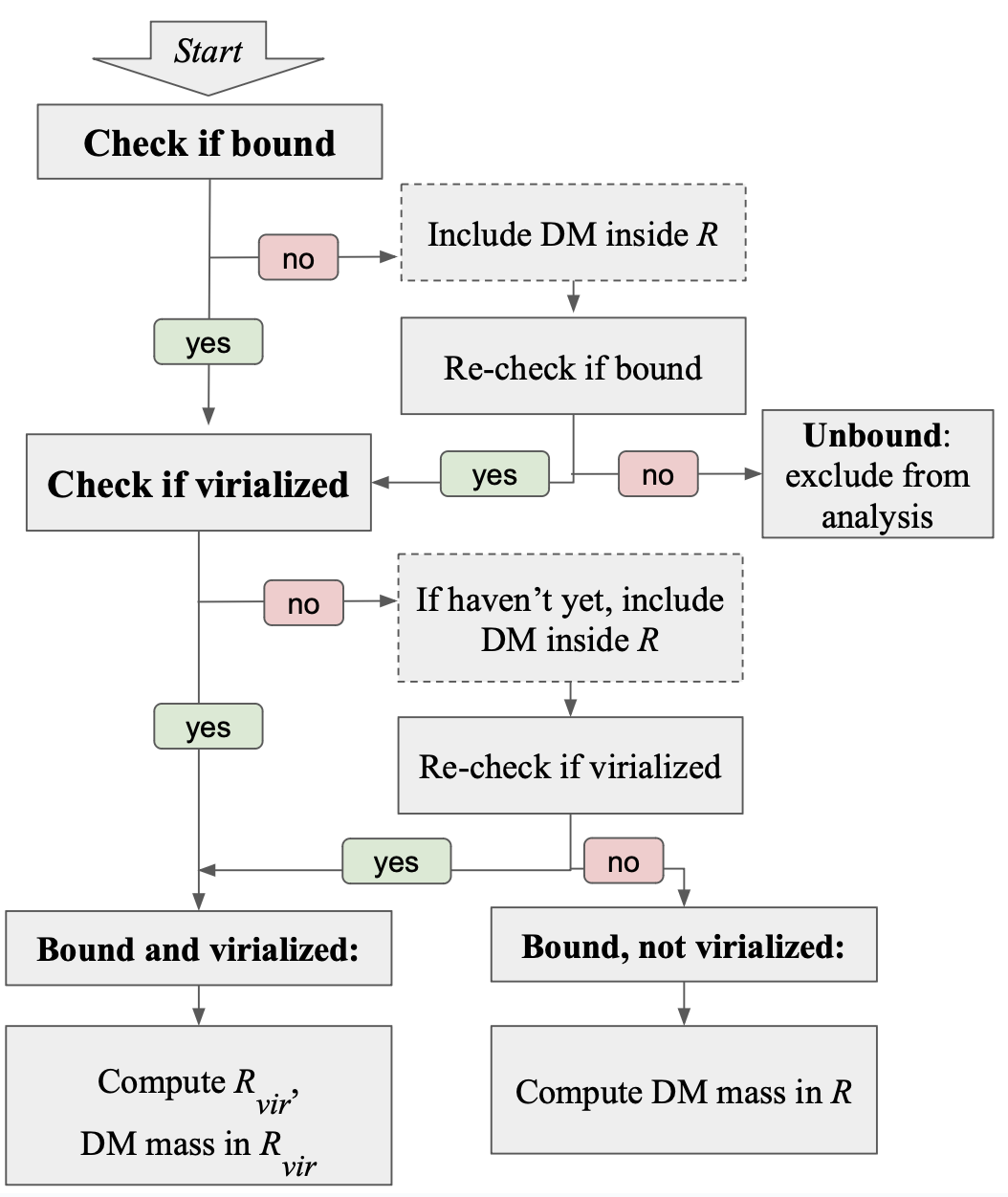}
    \caption{Flow chart schematic showing method for checking boundedness and virialization and computing DM mass within $R$.}
    \label{fig:flowchart}
\end{figure}

For bound structures, we check whether or not the system is virialized.
We use the criterion $1.5< -U/K< 2.5$ to determine virialization, where $U$ is the potential energy and $K$ is the kinetic energy. 
For those structures that we find to be virialized, we calculate the radius $R_{\rm vir}$. 
For this we take the radius for which $|U|/K=2.$
If a structure is not virialized, we characterize its radius as the $R_{\rm max}$, the galactocentric distance of the most distant star particle. 
This method is summarized in Fig.~\ref{fig:flowchart}, which depicts a flowchart of our process. 

\subsection{Association of DM halos}
In traditional algorithms, the DM halo mass is computed for the DM-primary structure using a common formalism, such as the mass within a radius $R_{200}$ corresponding to a region enclosing a density $200$ times the mean (or critical) density in the Universe. 
This process is used in our DM-centric runs, as it is straightforward to calculate $M_{200}$ within $R_{200}$ when DM is associated with the primary.
However, running a structure finding algorithm on the stars or gas particles/cells without associating DM as a primary or secondary particle complicates the process of calculating the halo mass. 
Here, we use two methods to understand the dark matter environment of our simulated structures. 
First, we search the halo catalogue provided by the DM-centric method, and for each star cluster identified by the baryon-centric algorithm, we find the nearest DM halo. 
However, we additionally seek to associate the DM particles that would be found ``observationally," near stars or gas observed electromagnetically. 
This may not correlate with the halos in the DM halo catalogue, as evidenced by Fig~\ref{fig:visuals}.
This limits us to the dark matter particles in the region of the baryonic particles. 
Thus, we also compute the mass of dark matter contained within a sphere centered at the center of mass of radius equal to that identified by the method of \S~\ref{sec:boundmethods} (either the virial or maximum radius, depending on the virialization state). 
In \S~\ref{sec:earlyclusterenv}, we compare these two probes of DM environment.

In the next sections, we have split our major results into three parts.
First, in \S~\ref{sec:boundclusters}, we compare the results of the baryon-centric and  DM-centric structure-finding methods. 
Then, in \S~\ref{sec:virializedclusters}, we describe the virialization state of the simulated structures using a simple analytic timescale model. 
Finally, in \S~\ref{sec:earlyclusterenv}, we use our methodology combined with the analytical understanding to paint a picture of early star clusters: their environment and properties. There, we analyze our systems in relation to the {\it JWST} star clusters with high surface density.

\begin{figure*}
    \centering
    \includegraphics[width = 0.45\linewidth]{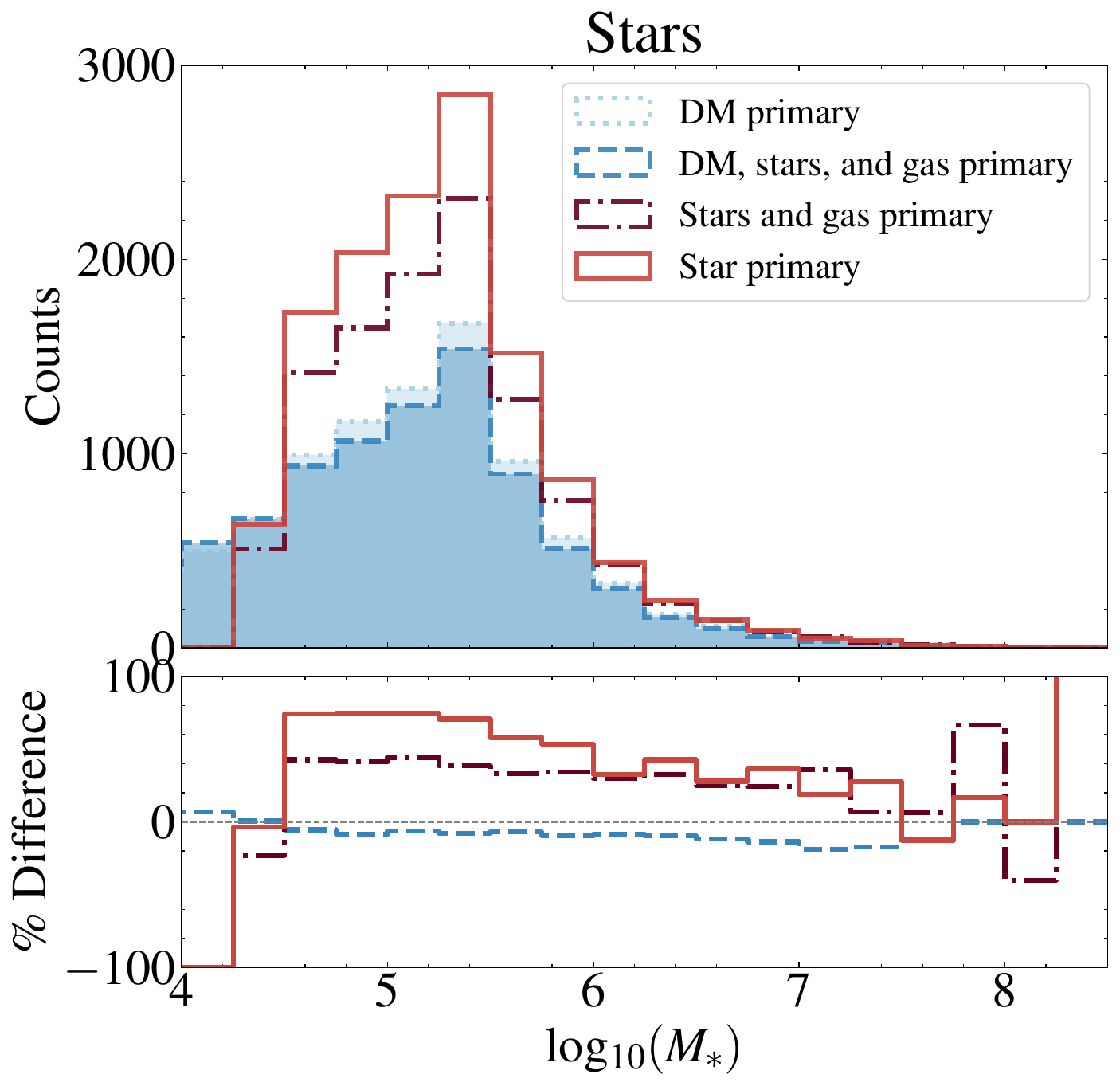}\includegraphics[width = 0.45\linewidth]{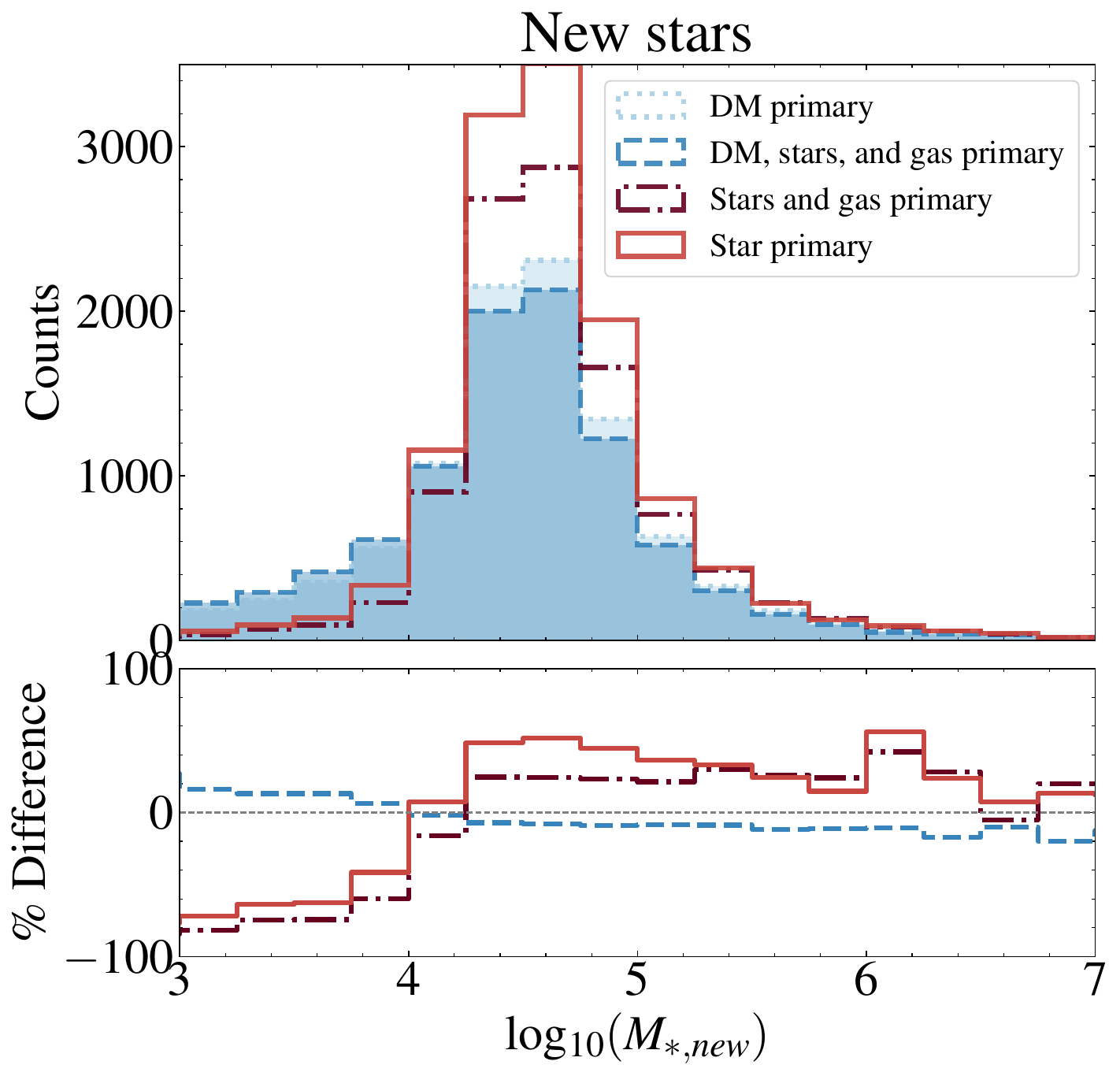}    
    \caption{Top left panel: Histogram of object counts by stellar mass at $z=12$ using algorithms centered on the particles listed in Table~\ref{tab:FOFruns}. Bottom left panel: Percent difference between each run and the standard DM-primary, stars$+$gas method ($f= N/N_{\rm DMP}$). The grey horizontal line shows $f=1$. 
    Blue shades are runs that include DM--the DM primary (dotted) and the stars, gas, and DM primary (dashed). The red shades are baryonic runs--the star primary (solid) and the stars and gas primary (dot-dashed). Top right panel: Same as the corresponding left panel, but objects are instead binned by newly-formed stellar mass at $z=12$. Bottom right panel: Percent difference between each run and the standard DM-primary, stars$+$gas secondary algorithm ($f= N/N_{\rm DMP}$). The grey horizontal line shows $f=1$. 
    }
    \label{fig:mstarhistz12}
\end{figure*}

\section{Comparison of algorithms targeting DM vs baryonic particles} 
\label{sec:boundclusters}
The left-hand panel of Fig.~\ref{fig:mstarhistz12} shows the number of galaxies binned by stellar mass found in our $z=12$ snapshot according to the different structure-finding algorithms. 
The bottom panel shows the fractional comparison with the standard DM-primary, stars$+$gas secondary algorithm. 
From the figure, it is clear that simply counting and binning the number of galaxies in a simulation snapshot depends strongly on the particle type used for the structure-finding algorithm. 
In Appendix~\ref{app:clustermassfunction}, we derive the star cluster mass function for the box from these number counts. 
Since this volume uses an increased $\sigma_8$ parameter, the provided mass function reflects that of  very high density peaks at this epoch and not the Universe on average. 

Variations in the number of galaxies of 50\% or more are present between the runs. 
This demonstrates that a system can form enough stars to represent a small galaxy by observational classification (i.e., its stars may be detected by a sufficiently sensitive observatory), but a dark-matter-focused algorithm may not find it. 
Alternatively, the dark matter primary algorithm may discover this structure but misclassify its stellar mass and star particles, since a merging system or a complex filamentary system may contain a stellar component at the edge.  
The DM$+$stars$+$gas primary (where baryonic and DM particles are combined as the primary particle type) was designed to combat this case and add weight to the systems with clustered baryonic components. However, as evidenced by Fig.~\ref{fig:mstarhistz12}, the DM$+$stars$+$gas primary number counts trace the dark matter only version to within a few percent at all masses. 

A relevant parameter for these galaxies is the number of newly formed stars in the snapshot, which is directly used to calculate the star formation rate \citep[e.g.,][]{Williams+24,Lake+24a}. 
We show histograms of the counts by newly formed stars ($M_{*,\rm new}$) in the right-hand panel of Fig.~\ref{fig:mstarhistz12}, and it is clear that the number at each mass varies significantly for this parameter as well. 
This is especially concerning given that the number counts by newly formed stars are used in the determination of the star formation rate and UVLF for these simulations\footnote{Without an indication of the underlying halo mass, it is difficult to accurately correct the runs for the increased clustering parameter $\sigma_8$ used in this study. 
So by this method, it is challenging
to predict the UVLF for the simulated box.}.  

To understand why the various structure-finding algorithms produce such number counts, let us consider again the visualizations of example systems in Fig.~\ref{fig:visuals}.
This Figure illustrates scenarios where the DM and baryonic methods disagree. 
In scenario (a), the central star cluster is obviously located at a density peak in the dark matter distribution. However, this object is not identified by the DM primary algorithm. 
Thus, a relatively massive star cluster/early galaxy is not counted by the DM-centric run.
In case (b), a star cluster is located near $R_{\rm 200}$ of a larger halo. It seems to be undergoing a merger and, again, is co-located with a DM density peak that is not identified as a halo. 
In cases where a merger is ongoing, the dark matter halo may be disrupted by a larger neighbor and thus is missed in the DM search. However, the star cluster remains and would be observationally detected by a theoretical observatory of sufficient sensitivity. 
In case (c), we see a highly complex region at the intersection of several filaments. In this case, it is clear that the approximation of spherical halos may not be appropriate. Here, the dark matter and gas in the region are located along filaments streaming into the central halo.
This could be observationally explored through studies of galaxy morphology \citep[e.g.,][]{pandya_galaxies_2024}.
In this complex region, star clusters have become spatially decoupled from their host halos, and thus, several clusters are found outside of the DM primary.

Note that in all three boxes, several star clusters are present that are either outside any dark matter halo (marked with an ``X") or located towards the edges of a central halo. 
Inside a halo, using the baryon-centric methodology allows us to identify individual clumps, which can give a more accurate accounting of the surface density. 
We will discuss the significance of the surface density in comparison to {\it JWST } observations in \S~\ref{sec:JWSTsurface density}.

Our analysis of these algorithms suggests that there is a population of relatively massive, bound star clusters that are hosted in halos that are undetected by the traditional dark matter based structure finding algorithm.
Potentially, they have halos too small to be considered by our threshold of 300 DM particles
or alternatively, the DM halos that host them do not meet other criteria of the halo-finding algorithm (such as being sufficiently dense). 
We note that simulations including the primordial cosmic streaming velocity between dark matter and baryons, which serves to separate gas from dark matter and generate more diffuse halos \citep[e.g.,][]{Lake+23b, Williams+23}, may suffer even more from this issue. 
Another group of star clusters may be in the process of accretion or fragmentation, and thus be missed as they straddle the boundary between the $R_{200}$ of multiple halos. 
\begin{table*}[]
\centering
    \begin{tabular}{|c|c|c|c|c|}
    \hline
         {\bf Object category } &$\mathbf{f_{\rm vir}(z=12)}$ &$\mathbf{f_{\rm vir}(z=13)}$&$\mathbf{f_{\rm vir}(z=14)}$&$\mathbf{f_{\rm vir}(z=14)}$  \\
         \hline \hline 
         DM-primary & 0.779  & 0.769 & 0.765 & 0.746 \\
         \hline
         Stars \& gas primary &  0.160 & 0.144 & 0.168 & 0.169 \\
         \hline
         Star primary & 0.174  & 0.158 & 0.156 & 0.153 \\
         \hline
    \end{tabular}
    \caption{Virial fractions in each run at output  redshifts. 
    Results include all halos with $\log_{10}{(M_*)}>4.25$.}
    \label{tab:virial_fraction}
\end{table*}

\begin{figure}%
    \centering
    \includegraphics[width = 0.95\linewidth]{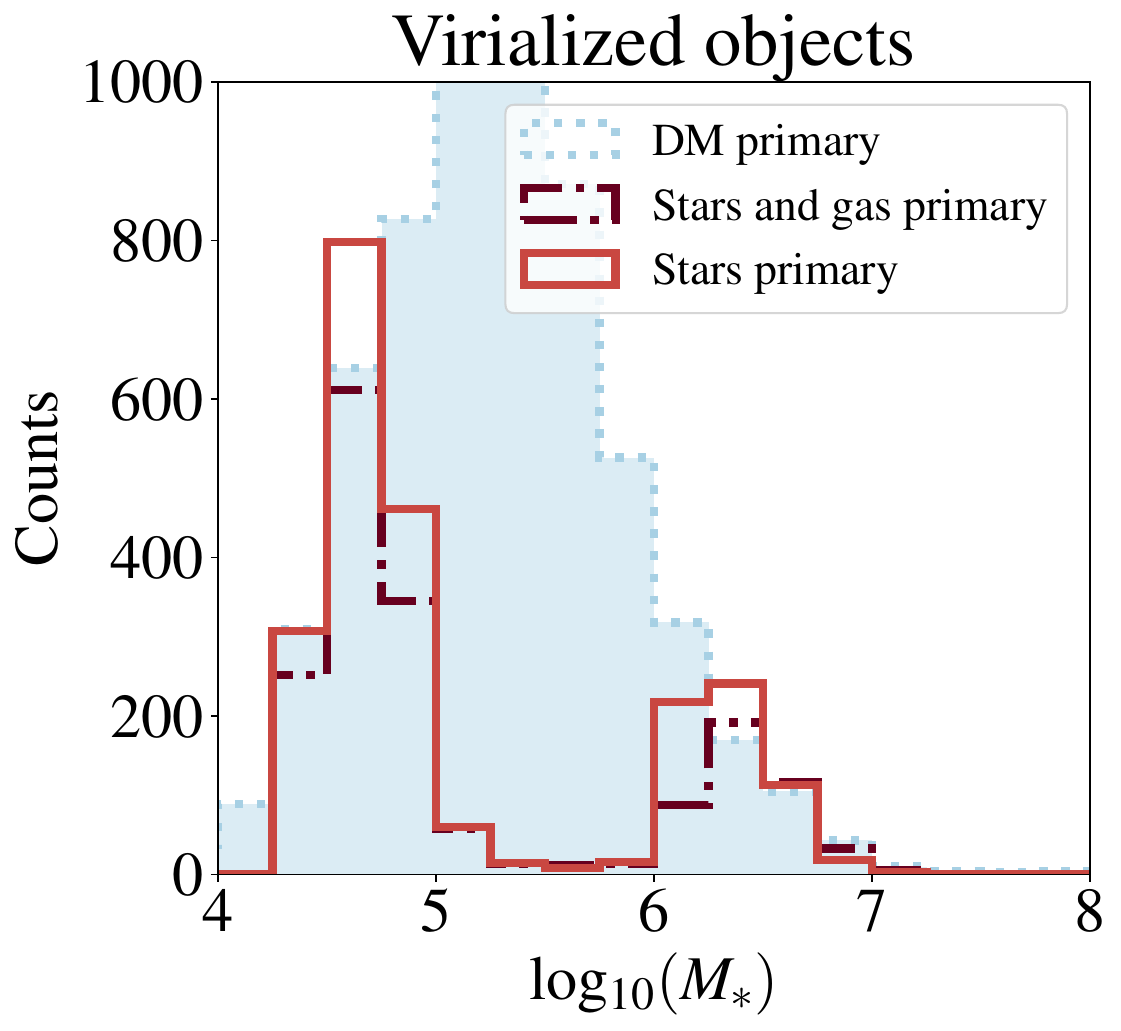}
    \caption{
    Number of virialized objects at $z=12$ by stellar mass in solar masses. 
    Here, we consider the star primary (red solid line), star+ gas primary (dark red dot-dashed line), and DM primary (light blue dotted line).
    }%
\label{fig:fvir}%
\end{figure}

\section{Dynamical properties of star clusters}
\label{sec:virializedclusters}

In isolation, we expect that long-lived star clusters will tend towards equilibrium. 
In this Section, we investigate the virial state of star clusters in our simulation. 
We use simple dynamical arguments to estimate the timescales for virialization at various masses and use this to understand the simulated structures. 

\subsection{Distribution of virialized systems}

As described in \S~\ref{sec:methods}, for runs with only baryonic (stellar or gas) particles as the primary particle type, one concern is that the structure-finding algorithm may pick up unbound structures that are numerical artifacts. 
As detailed in that section, we checked all objects for boundedness based on their total energy. 
During that process, we also tested whether or not systems are virialized. 
We find that even after including DM particles in the immediate region of the stars, the majority of objects are not virialized. 
Table~\ref{tab:virial_fraction} top panel shows the virial fraction in each run that is, the number of virialized objects ($N_{\rm vir}$) per total number ($N_{\rm tot}$):
\begin{equation}
    f_{\rm vir} =\frac{ N_{\rm vir}}{N_{\rm tot}}.
\end{equation}
Around 80\% of systems detected through the DM primary are virialized.
However, as shown in the Table, for baryon primary objects, the virial fraction varies slightly but remains around 16\% for all redshifts. 
While systems are theoretically expected to tend toward equilibrium over time, the process of virialization may be disrupted by external factors, such as mergers, accretion, and tides. 
The fact that a greater fraction of the baryonic-centric systems are not virialized hints that the systems that are not detected through the DM primary algorithm are systems undergoing disruptive processes, such as mergers, which additionally cause non-detection of the host halo. 

To investigate this behavior, in Fig.~\ref{fig:fvir}, we plot the distribution of virialized objects by their stellar mass at $z=12$.  
For the DM-centric detected structures (blue histogram), the distribution is similar to the overall distribution of halos as in Fig.~\ref{fig:mstarhistz12}. 
Indeed, for halos with $M_*>10^5 M_\odot$, a Kolmogorov-Smirnov test cannot reject the null hypothesis that the virialized and overall distributions are the same ($p-$values between $0.17-0.54$ for $12\leq z\leq 15$). 
In other words, a roughly constant fraction of objects at most masses are virialized. 
However, the baryonic-centered runs (shown in red in the Figure) show a bimodal distribution, with a dearth of virialized objects between $10^5M_\odot$ and $10^6 M_\odot$ 
\footnote{For these virialized objects, the null hypothesis that they arise from the same distribution as the overall population fails with $p-$values below $10^{-120}$ for  $12\leq z\leq 15$.}.
This bimodality is found in every redshift where a snapshot was taken in the simulation (see Appendix~\ref{app:redshifts}). 
Since the overall halos (traced by the dark matter) are virialized, this suggests processes affecting the conversion of gas to stars or the stellar systems themselves. 
The following two Subsections investigate these processes in detail. 

\subsection{Early gas conditions}
\label{sec:gasconditions}

\begin{figure}
    \centering
    
    \includegraphics[width = 0.95 \linewidth]{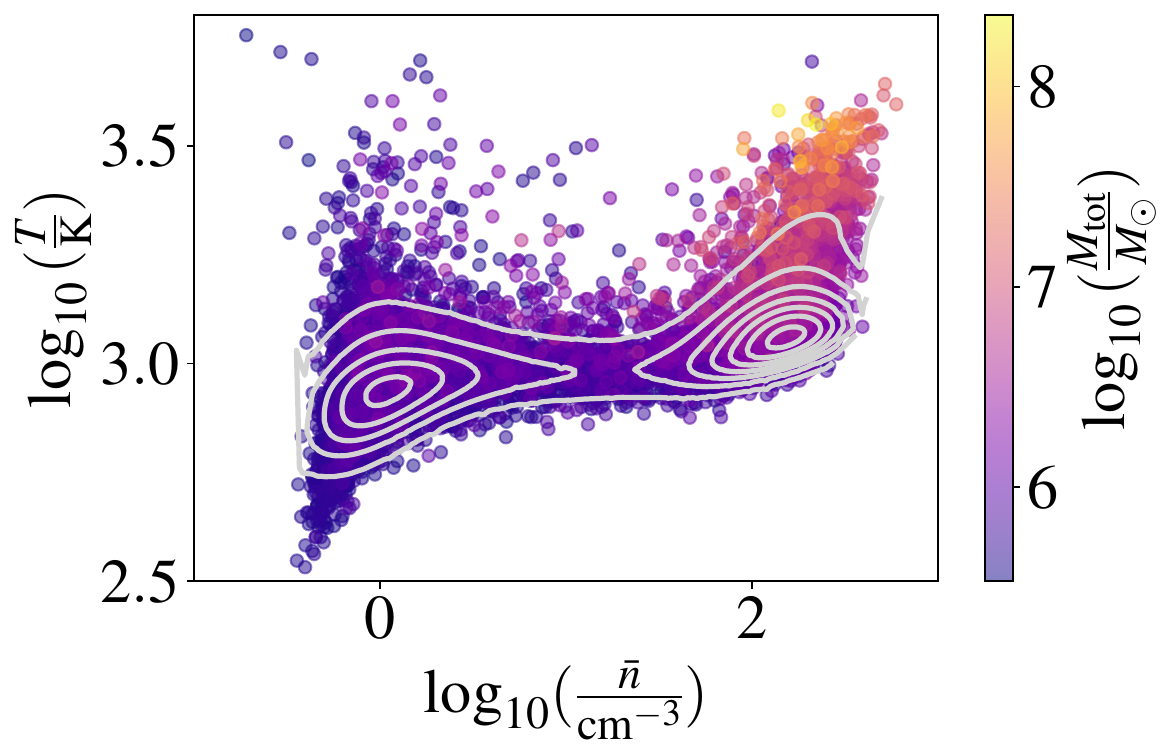}
    \includegraphics[width = 0.94\linewidth]{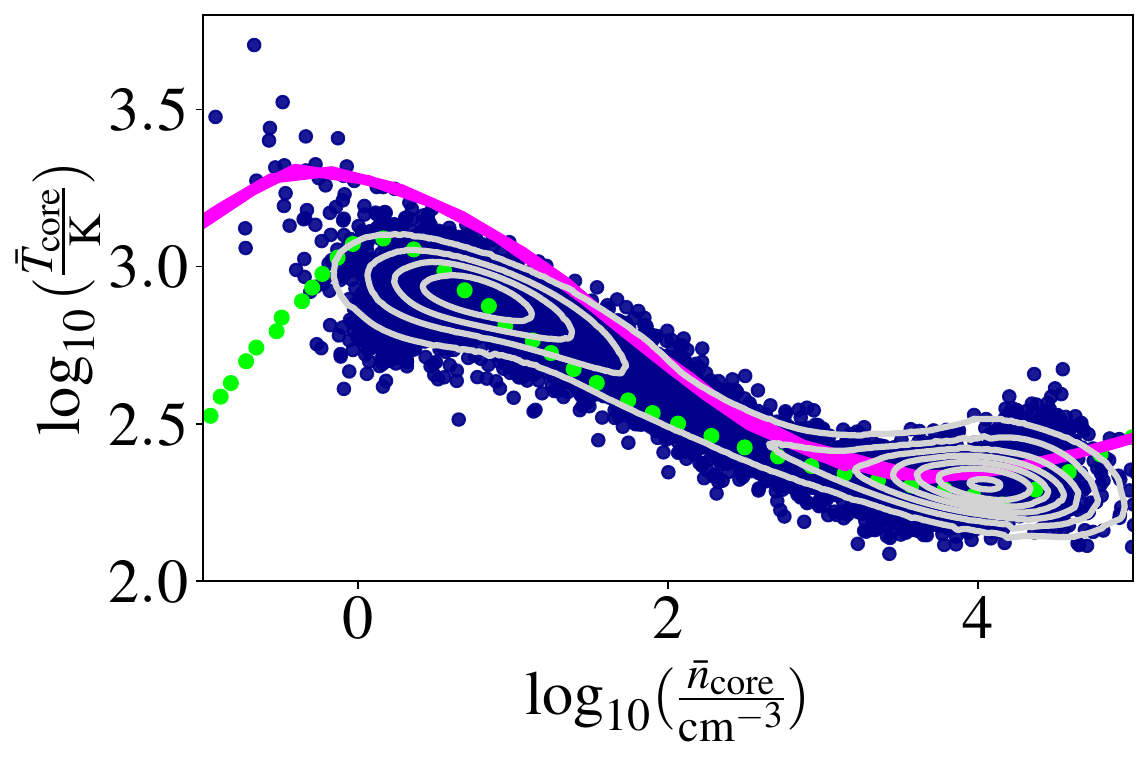}
    \caption{Top panel: Logarithm of average temperature versus average number density at $z=25$ in DM halos.
    Contours show the 2D density of points in the parameter space. The color bar depicts total halo mass ($M_{\rm DM}+M_{\rm b}$). 
    Bottom panel: Same as top panel, but the relation is plotted for the core gas cell in each halo instead of the mean. 
    Overplotted are the results of \cite{omukai_low-metallicity_2010} (green dotted) and \cite{Yoshida+06} (pink solid).
    }
    \label{fig:temp-density}
\end{figure}

To uncover the origin of the two populations of virialized objects, 
we start by considering the state of the gas at slightly higher redshift. 
This baryonic matter is the material that must condense and cool to form the star clusters seen at our final redshifts. 
We take a snapshot from the simulation at $z=25$, and run only the standard DM-primary, stars$+$gas secondary algorithm. 
For these early halos, we plot the average gas temperature versus average number density in Fig.~\ref{fig:temp-density}.
Here, we see that the halos fall into two typical densities, roughly independent of mass and temperature at low masses. 
 These correspond with typical molecular cooling tracks as seen in other works \citep[e.g.,][]{Yoshida+06,greif_first_2010,omukai_low-metallicity_2010,Hirano+18,Lake+24b}.
This is shown by the bottom panel of Fig.~\ref{fig:temp-density}, which compares the density and temperature in the halo core to other works: there is good agreement and the two populations correspond to a buildup along the cooling tracks.
In other words, primordial molecular hydrogen cooling seeds two characteristic densities into the early gas, according to each halo's evolution along the cooling track.

We postulate that the two gas densities present already in the DM halos at $z=25$ may relate to the two virialized populations we find at lower redshift.
In the following section, we estimate two characteristic densities for an analytical investigation corresponding to the peak density of each of these two populations. 

\subsection{Timescale investigation}
\label{sec:timescales}
\begin{table}[]
    \begin{tabular}{|c|c|c|}
         \hline {\bf Parameter} & {\bf Value} \\
         \hline
         \hline
        $\rho_1$ & $3.16  ^{+6.83}_{-2.16}\times 10^{-22} \text{ g cm}^{-3}$ \\
        \hline 
        $\rho_2 $ & $ 1.58_{-3.43}^{+0.954} \times 10^{-24} \text{ g cm}^{-3}$\\
        \hline
         $\bar{v}_{\rm group}$ & 26.3 km s$^{-1}$ \\
         \hline
         ${\sigma}_{\rm group}$ & 16.6 km s$^{-1}$ \\
        \hline
        $\bar{N}_{\rm DM}$ & $6.06\pm 3.89$ \\
        \hline
    \end{tabular}
    \caption{Derived quantities from simulated results used in analytical calculations of timescales. Further data is presented in Appendix~\ref{app:simulatedquantities}.}
    \label{tab:derivedvalues}
\end{table}
\begin{figure*}
    \centering
    \includegraphics[width=0.7\textwidth]{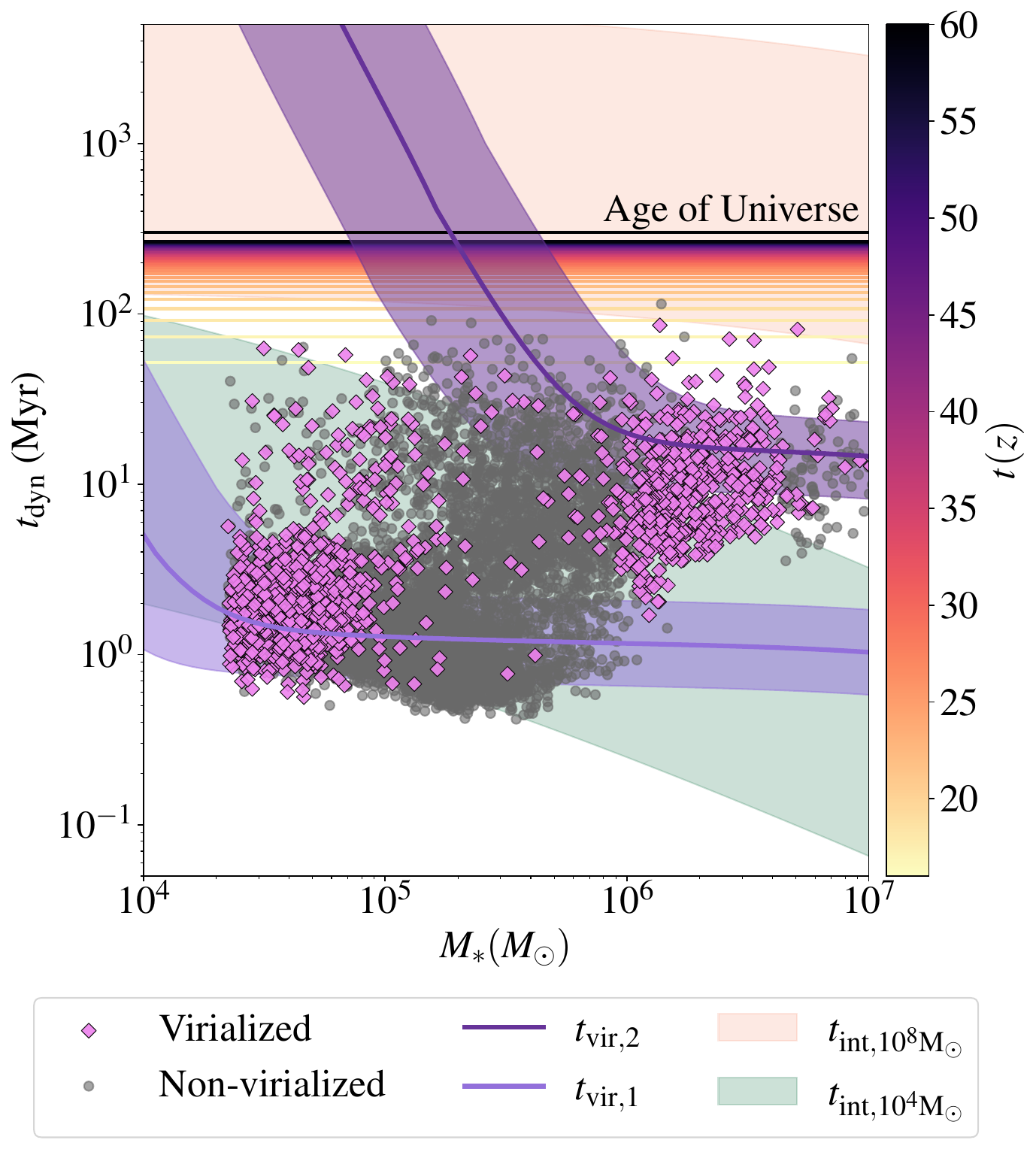}
    \caption{Analytic virialization and interaction timescales compared to dynamical timescales of Star+Gas primary objects at $z=14$.
    $t_{\rm vir,1}$ is the virialization timescale for systems with density $\rho_{1}$, and $t_{\rm vir,2}$ corresponds to $\rho_{2}$ (see Table~\ref{tab:derivedvalues}). Shaded regions around these curves denote the range of values associated with the simulated parameters.
    The dynamical timescale is calculated from the star particles' orbital velocity. Similar results are obtained if the gas density or gas cell orbital velocities are used to trace the dynamical timescale. The color bar shows the time since a variety of redshifts in Myr. 
    The solid black line shows the age of the universe at $z=14. $ The red region shows the range of interaction timescales for a $10^8M_\odot$ halo, and the green region shows the interaction range of timescales for a $10^4 M_\odot$ halo. }
    \label{fig:data-tscales}
\end{figure*}
Given the two characteristic densities present in halos from high redshift in the simulation, we understand the state of virialization of simulated objects in the context of a simple timescale argument. 
Specifically, in Fig.~\ref{fig:data-tscales}, we consider each star cluster dynamical timescale $t_{\rm dyn} = r/v_c$, where $r$ is the cluster radius and $v_c$ is the mean circular velocity of the star particles\footnote{The dynamical timescale was calculated using the star particles' orbital velocity. We repeated this analysis with other tracers of the dynamical time, such as gas density and gas cell velocity, and similar results were obtained.}.
This time scale is plotted as a function of mass. 
As depicted, we can identify two branches of virialized systems associated with two separate mass distributions. 
We find the densities corresponding to the two peaks of the average density distribution for this analysis (see Fig.~\ref{fig:temp-density} top panel). 
These are
$n_1 = 1.89\times 10^2 \text{  cm}^{-3}$ and 
$n_2 =9.48\times 10^{-1} \text{  cm}^{-3}$. 
In terms of mass  density, $\rho_1 = 3.16 \times 10^{-22} \text{ g cm}^{-3}$ and  $\rho_2 = 1.58 \times 10^{-24} \text{ g cm}^{-3}$.

Then, to understand this behavior we consider the dynamical processes governing the clusters' evolution.

Our analytical method to uncover the underlying dynamics is to compare the minimum timescale for virialization to the range of typical timescales for interaction with nearby halos. 
The minimum timescale for virialization is a function of mass and density: $t_{\rm vir} (M,\rho)$ (see below for a  detailed expression).  
The timescale for interaction is additionally a function of the mass of the nearby halo: $t_{\rm int}(M,\rho, M_{\rm DM})$ (again, derived below). 
There are three relevant scenarios: 
\begin{enumerate}
    \item $t_{\rm vir}(M,\rho,)>t_{\rm Universe}$: If the virialization timescale is longer than the age of the Universe (the simulation run time), we do not expect an object to be able to virialize in our simulation. 
    \item $t_{\rm int}(M,\rho, M_{\rm DM})<t_{\rm vir} (M,\rho)<t_{\rm Universe}$: If interactions with nearby halos occur much more frequently than the virialization timescale, we do not expect virialization, as clusters may be disrupted or merged with nearby objects before they reach equilibrium. 
    
    \item $t_{\rm vir} (M,\rho)<t_{\rm int}(M,\rho, M_{\rm DM})<t_{\rm Universe}$: If the interaction timescale is longer than the virialization timescale, then the system has time to virialize before being disrupted again, and we expect to find it in a state of virialization.
\end{enumerate}
These timescales are used to explain the bifurcation of the clusters' virialization properties and are derived in more detail below. We first describe the behavior followed by the relevant equation. 

The minimum virialization timescale is plotted in purple for our two characteristic densities in Fig.~\ref{fig:data-tscales}. 
Additionally, we show a range of reasonable values for the interaction timescale in the orange and green shaded regions (there is a large range of  environments in the simulation).
The green region corresponds to halos of only $M_{\rm DM}\sim10^4 M_\odot$. 
The orange shaded region corresponds to halos of $M_{\rm DM}\sim10^8 M_\odot$ 
(see Fig.~\ref{fig:analyticaltscales} in the Appendix for more values). 
We show the age of the Universe as a horizontal black line, as well as the time since various redshifts as horizontal lines corresponding to the values shown on the color bar. 
The bifurcation in dynamical timescales of the simulated points corresponds to the two populations by density (see Fig.~\ref{fig:rho_m} in Appendix~\ref{app:simulatedquantities}), and the data points lie roughly along the two tracks for analytic virialization timescales. 

From this, we can use the  criteria above to correctly predict whether a population at a given mass will be virialized. We note that the correct interaction timescale is that corresponding to a halo mass that would cause a major merger (
$\sim > 10\% M_*$). 
Firstly, our model correctly shows that low density objects ($\rho_2, t_{\rm vir, 2}$) should not be virialized below $M_* \sim 10^6 M_\odot$. 
However, those that do virialize should stay virialized, since the interaction time for very massive halos is much longer than their virialization time. 
For high density objects, our model also explains the mix of virialized and non-virialized systems. 
For example, a roughly $10^4 M_\odot$ cluster of high density should very frequently be virialized, since the interaction timescales (even for similarly low mass halos) are typically longer. 
Meanwhile, a $10^{5.5}M_\odot$ cluster may often be found in a disrupted state, since the interaction timescale for halos $10-100\%$ of its size dips below its virialization timescale.

\paragraph{Derivation of the virialization timescale}
\label{sec:vir_derive}

Here, we describe in detail how we analytically estimate the minimum timescale for star cluster virialization (the purple lines in Fig.~\ref{fig:data-tscales}). 
To do this, we assume that the baryonic matter must undergo two stages: first, gas must cool such that gravity may overcome gas pressure and form stars. Second, the stars must undergo violent relaxation, such that virialization occurs. 
Inspired by the methods of \cite{Lake+22}, we estimate the timescale for molecular cooling with a two phase model: first, the gas cools isochorically, until the Jeans mass falls below the total mass. Then, the gas collapses to form stars on the cooling timescale \citep[defined in ][]{Schauer+21}. 
In this model, our cooling timescale is:
\begin{equation}
    t_{\rm cool} = \frac{k_{\rm B} T_{{\rm collapse}}}{n_{\rm H} (\gamma-1)\Lambda(T_{{\rm collapse}})f_{\rm H2}} +t_{\rm iso} \ ,
    \label{eq:cooling}
\end{equation}
where $T_{\rm collapse}$ is the temperature of collapse after the Jeans mass is exceeded, $n_{\rm H} $ is the  hydrogen number density, $f_{\rm H2}$ is the molecular hydrogen fraction, $\gamma$ is the adiabatic constant, $k_{\rm B}$ is the Boltzmann constant, and 
$\Lambda(T)$ is the temperature and density dependent molecular cooling rate. We use the rates provided by \cite{GP+98}:
\begin{equation}
    \Lambda(T) = \frac{\Lambda(LTE)}{1+[n^{\rm cr}/n({\rm H})]} \ . 
\end{equation}
Again following \citet{Lake+22}, the initial temperature is taken to be 500 K or the Jeans temperature, whichever is higher:
\begin{equation}
    T_{\rm J} = \frac{3 {\rm G} \mu m_{\rm H} }{5 k_{\rm B}}\left(\frac{36M^2 \rho}{\pi^2}\right)^{\frac{1}{3}} \ . 
\end{equation}
If the Jeans mass is not initially reached, the isochoric cooling timescale in Eq.~(\ref{eq:cooling}) is:
\begin{equation}
    t_{\rm iso} = \int_{T_{\rm Collapse}}^{T_{\rm init}}\frac{{\rm k}_{\rm B}}{n_{\rm H} (\gamma-1)\Lambda(T)f_{\rm H2}}dT \ .
\end{equation}
The cooling timescale is plotted in blue in the Appendix~\ref{app:analytictimescales}~Fig.~\ref{fig:analyticaltscales} for $\rho_1$ and $\rho_2$. 
For high densities, the cooling timescale on the order of Myr or well below for most of the masses considered. 
For the lower density case, the cooling timescale exceeds the age of the universe at these redshifts for masses below $\sim 10^5$ M$_\odot$. 

Once a star cluster has formed, the next phase of virialization is violent relaxation. 
This requires a rapidly changing gravitational potential, such that the orbits of stars experience significant perturbation on dynamical timescales.  
A detailed estimate would require an understanding of the complex environment surrounding each star cluster. 
Here, we seek to instead provide a minimum timescale for virialization, in order to understand whether or not we should expect many clusters to virialize at each mass and density. 
In other words, this estimate assumes that the star cluster undergoes virialization immediately following its formation. 
To place this lower bound, $t_{\rm vir},$ our ``virialization timescale" is simply:
\begin{equation}
    t_{\rm vir} = t_{\rm cool} + t_{\rm dyn} \ .
    \label{eq:virialization}
\end{equation}
This sum is what is plotted in purple in Fig.~\ref{fig:data-tscales}.
We estimate the dynamical timescale for  mass $M$ and density $\rho$ by assuming a uniform density. 
In this case,
\begin{equation}
    t_{\rm dyn}= \frac{r}{v_c} \ ,
\end{equation}
where $v_c$ is the circular velocity of a halo of constant density with mass $M_{\rm tot} = M_{\rm g}  + M_*+ M_{\rm DM}$. 
For this estimate, we numerically fit $f_{\rm g} = M_*/M_{\rm g}$ and $M_*/M_{\rm DM}$ to match our simulations (see Appendix~\ref{app:simulatedquantities}).
This lends the dynamical timescale a very weak dependence on mass. 
In Appendix~\ref{app:analytictimescales}~Fig.~\ref{fig:analyticaltscales}, the dynamical timescales for $\rho_1$ and $\rho_2$ are shown in grey. 

The overall $t_{\rm vir}$, (Eq.~(\ref{eq:virialization})), is plotted in Fig.~\ref{fig:data-tscales} for both densities as a purple line. 
For low masses, the virialization timescale is dominated by the gas cooling, whereas for high masses, the timescale is set by the dynamical time. 
Already, without including any disruption processes, we see from the two purple lines in Appendix Fig.~\ref{fig:analyticaltscales} that the low-density gas should not virialize within the age of the Universe for systems below $M_*\sim 10^{5.5} M_\odot$. 
This aligns well with the trough of virialization in Fig.~\ref{fig:fvir}.
However, this does not explain why high density systems, if they exist for higher masses, do not achieve virialization for all masses.
For this reason, we invoke the interaction timescale, derived below.

\paragraph{Derivation of the interaction timescale}
\label{sec:int_derive}

The most important process disrupting virialization for the purposes of our estimation here is the interaction with the surrounding environment\footnote{Tidal heating was also investigated as a process that may possibly disrupt the virialization state of a cluster. For the systems considered in this paper, the diffusive tidal heating timescale was several orders of magnitude longer than the age of the Universe at this epoch.}. 
As our simulations are designed to follow the first epoch of galaxy formation, when the hierarchical merger mechanism is building up larger galaxies, we expect interactions to be quite common. 
Accretion of external star clusters may cause virialized clusters to move out of equilibrium. 
However, the degree and timescale of disruption depends on the mass ratio of the interaction. 
Major mergers, or interactions with other structures of similar mass, should be expected to severely impact the virial status of a system \citep{Yoshida03}.
Meanwhile, accretion of a halo whose mass is orders of magnitude smaller may not be enough to significantly disrupt equilibrium. 
We note that since this is a gravitational process, all halos (not just star cluster hosts) should be included in the calculation.

A simple rate calculation provides the timescale for a nearby halo to enter within the virial radius of the dark matter halo hosting the star cluster in question: 
\begin{equation}
    t_{\rm int} \sim \frac{1}{n_{\rm DM}(M)\sigma_{\rm int} v_{\rm group}} \ .
    \label{eq:tinteract}
\end{equation}
Here, $n_{\rm DM}(M)$ is the number density of dark matter halos of mass $M$ in the region, $\sigma_{\rm int}$ is the cross section of interaction, $\sigma_{\rm int} = \pi (r_{\rm group}+r_{\rm perturber})^2$ and $v_{\rm group}$ is the group velocity of the star cluster with respect to the bulk flow. 
We estimate the radii in the cross-section $\sigma_{\rm int}$ using the analytical formula for virial radii as a function of mass and redshift from \cite{BarkanaLoeb+01}:
\begin{multline}
    r_{\rm vir} = 0.784 \left(\frac{M}{10^8 h^{-1} M_\odot}\right)^{1/3} \left(\frac{\Omega_m \Delta_c}{\Omega_m^z 18\pi^2}\right)^{-1/3} \\ \times \left(\frac{1+z}{10}\right)^{-1}h^{-1} \text{ kpc} \ .
\end{multline}

The other quantities in Eq.~\ref{eq:tinteract} are estimated from a reasonable numerical range in the simulations. 
For the number density of halos, we provide an upper and lower bound using two numerical estimates drawn from our simulations. 
Appendix~\ref{app:simulatedquantities}~Figure~\ref{fig:numberDM} shows the mean number of dark matter halos within a radius of 10 ckpc of each star primary object.
For most low-mass objects, there are around 6-7 dark matter halos in the vicinity. Assuming these are evenly distributed, we use the mean $N=6$ (see Table~\ref{tab:derivedvalues})  to estimate the number density in Eq.~(\ref{eq:tinteract}) for an interaction with the local environment. 
This figure likely overestimates the interaction time because it assumes that the nearby halos are uniformly distributed in the region. 
However, we know that structure is highly clustered, and often many halos are located along a filament of the cosmic web (see Fig.~\ref{fig:visuals}). 
Thus, as a lower bound, we estimate the density from the nearest dark matter halo, calculating the timescale for that interaction. 
This ``nearest neighbor" estimate provides an upper limit. 
It is most accurate for low-mass halos, which have the highest overall number density. 
Thus, structures are most likely to have a low mass halo as their nearest neighbor. 
Figure~\ref{fig:velocityhist} in Appendix~\ref{app:simulatedquantities} shows histograms of the mass of the nearest neighbor, velocity with respect to the bulk flow, and the number of nearby neighbors. 
We find the average group velocity of the star clusters to be $26 \text{ km s}^{-1}$ with respect to the bulk flow, with a standard deviation of $16.6 $ km s$^{-1}$.
We use the standard deviation to represent the typical relative velocity difference between two nearby neighbors. 

Finally, we must modify the number density to account for the variation of number density with mass. 
To do this, we use the \cite{Sheth+99} number density, and compute: 
\begin{equation}
    n(M)= f_{\rm \delta M}(M)n_{\rm est, sim},
    \label{eq:numberdens}
\end{equation}
where $f_{\rm \delta M}$ is the fraction of halos at mass M,
\begin{equation}
    f_{\rm \delta M}(M)=\frac{N(>(M+\delta M))-N(>M)}{N_{\rm tot}}.
\end{equation}

Using these quantities, we plot  $t_{\rm int}$ in Appendix~\ref{app:analytictimescales} Fig.~\ref{fig:analyticaltscales}. 
The left panel of the plot shows the upper bound interaction timescale with the local environment as a series of lines colored by interacting halo mass. 
The inverse dependence on mass reflects the larger cross section of larger halos, while larger interacting halo mass leads to longer timescales because of their lower number density. 
In the right panel of the Figure, the lower bound nearest neighbor timescale is shown, also colored by interacting halo mass. 
These timescales are much shorter as the nearest halo is much closer than the average halo.

\section{Early star clusters}
\label{sec:earlyclusterenv}
Now, we combine our methodology that identifies dense clusters of stars with the numerical and analytical understanding of the processes governing equilibrium to provide an overall explanation of the nature of cosmic dawn star clusters in our simulation box. 

\subsection{Environment}
In the previous sections, we have shown that the population of low-mass clusters and galaxies deviates from a simple theoretical model where clusters of stars reside in spherical or ellipsoidal halos. 
From our analysis, it seems that the role of environment in shaping their distribution is central. 
In this section, we further investigate the population in our box with respect to their local environment. 

First, we plot the stellar mass of baryon-centric-detected objects as a function of DM mass enclosed within their radius $R$ (see \S~\ref{sec:methods}). 
\begin{figure}%
    \centering
    \includegraphics[width = 0.45 \textwidth]{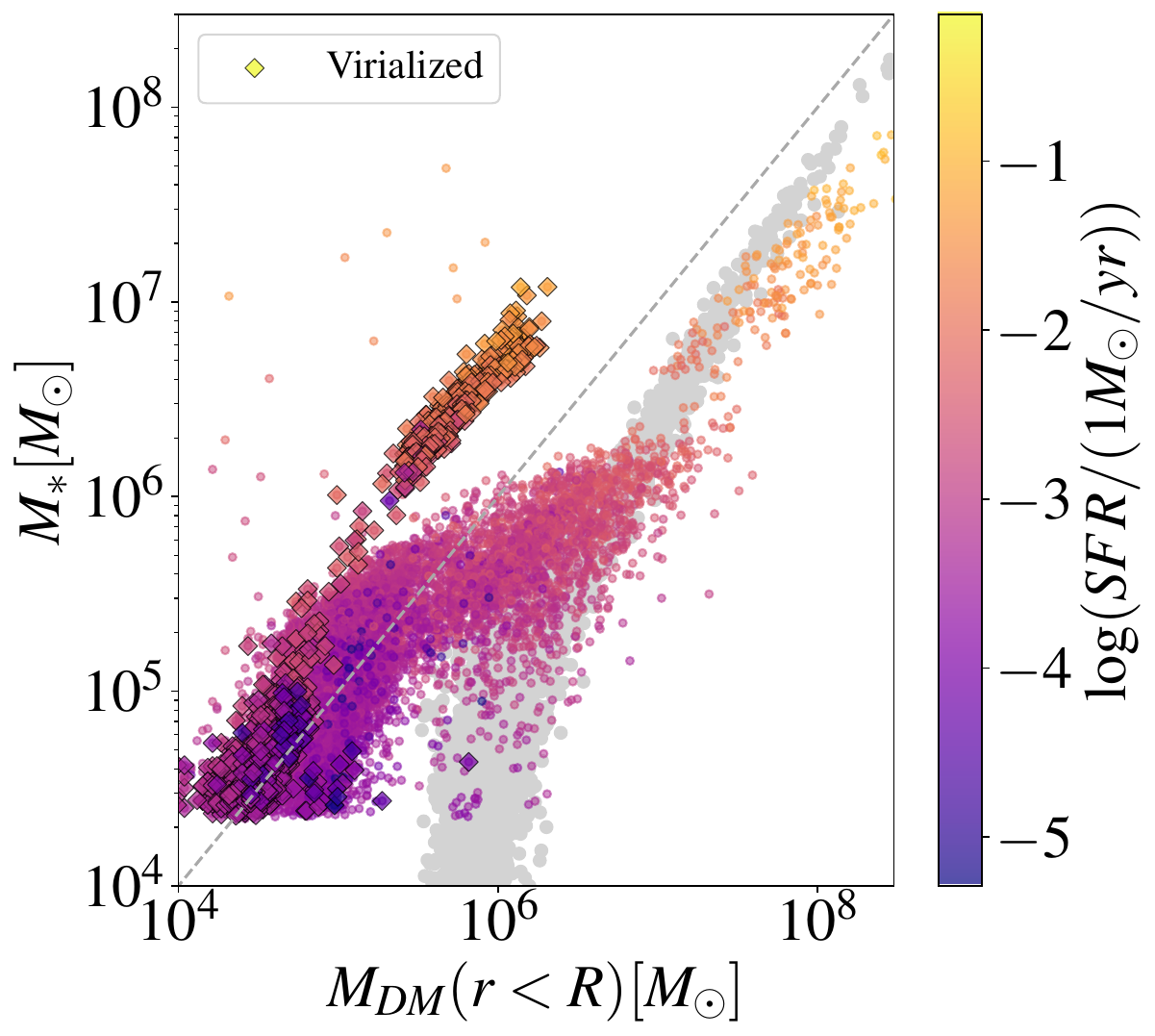}
    \includegraphics[width = 0.45 \textwidth]{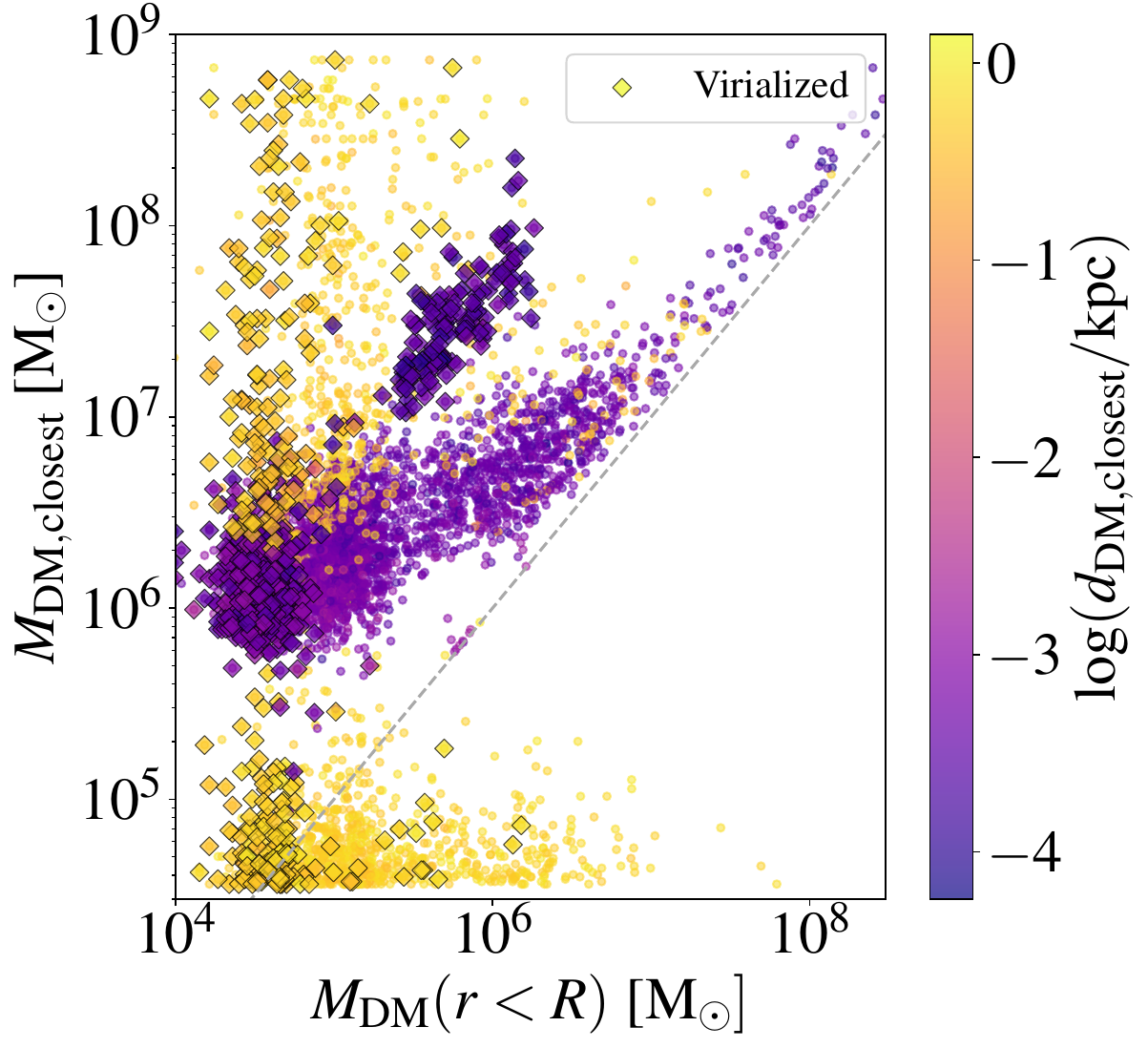}    
    \caption{Top panel: Stellar mass vs. dark matter mass within $R_{\rm obj}$ at $z=12$. $R_{\rm obj}$ is defined as the radius given from the FOF algorithm or the virial radius, if the object is virialized (see \S~\ref{sec:methods}). The color bar gives the logarithm of the star formation rate (SFR) Points shown are for the gas + stars primary. Bottom panel: Mass of the closest dark matter halo in the DM FOF runs vs. the enclosed DM mass within the object radius for the star + gas primary run at $z=12$. Points are colored by the logarithm of the distance to the closest halo.
    In both panels, the grey dashed line shows the $1:1$ relation.
    }%
    \label{fig:mstar-mdm}%
\end{figure}
Virialized objects emerge as a track with higher ${M_*}/{M_{\rm DM}(r<R)}$ 
ratio than the non-virialized population. 
Since the radius considered in this calculation for virialized objects is the virial radius $R_{\rm vir}$, which may be considerably smaller than $R_{\rm max}$, it follows that less dark matter is  enclosed. 
The two tracks converge at low masses, suggesting that $R_{\rm vir}\sim R_{\rm max}$ for these structures. 
However, at larger masses, the divergence in Fig.~\ref{fig:mstar-mdm} suggests that objects have a diffuse stellar component at greater distances than the virial radius of the central cluster.
Additionally, underplotted in grey are the relations from the DM-centric structure catalog. Generally, these converge with the baryon-centric run for the intermediate (and high) masses for the non virialized case, suggesting that the radius of virialized objects is similar to the dark matter $R_{200}$ at those masses. 

Similar trends are present in the lower panel of Fig.~\ref{fig:mstar-mdm}, which compares the DM mass of the nearest halo from the dark matter-centric run to the dark matter mass within the radius of the stellar object.
Many non-virialized objects lie along the $1:1$ curve, again suggesting that the closest dark matter halo is indeed the host halo of the object, and that the stars in non-virialized systems represent a more diffused component of similar radius to the dark matter. 
The virialized objects lie mostly above the $1:1$ line, suggesting that they have tight virial radii that do not align with the DM $R_{200}$. 
From the colorbar, which denotes the distance to the nearest dark matter halo, we can see that these virialized structures are indeed co-located with a dark matter halo, but in the enclosed mass measurement we are only tracing its core. 

Another reason structures may be deficient in dark matter is that they represent a fragmented or recently accreted system, and thus they lie at some distance from the central dark matter cusp. 
The colorbar indicates these objects lie at a large distance from the nearest halo in the DM catalog.

In some cases, objects lie below the $1:1$ curve. 
Nearly all of these systems are located at a great distance from the nearest halo (colored yellow), compared to the hypothetical separation from a host halo, which would lie nearly on top of the star cluster. 
However, their location on the $x-$axis shows that there is a large mass of dark matter in the region. 
This reveals the population mentioned previously in \S~\ref{sec:methods}: star clusters that live in halos not identified by the DM-centric algorithm, which would be missed by a DM-primary  only study. 
Of course, these may coincidentally lie elsewhere in the plot as well. 
From our visualizations in Fig.~\ref{fig:visuals}, we can see that this is especially frequent for $virialized$ objects, which are typically found in the core of halos--the dark matter-centric run has not picked up their host halo. 
Based on this visual analysis, we postulate that almost all the virialized objects that appear to have a large distance to their nearest halo instead are hosted by non-detected halos. 
This emphasizes why a subhalo algorithm method -- such as using {\tt AREPO}'s built in subhalo flag (which uses the SubFind algorithm) to identify fragmented clusters -- might still be unable to identify all the star clusters in the simulation box.

\subsection{Comparison with JWST observations}
\label{sec:JWSTsurface density}
\begin{figure*}
    \centering
    
    \includegraphics[width = 0.9\textwidth]{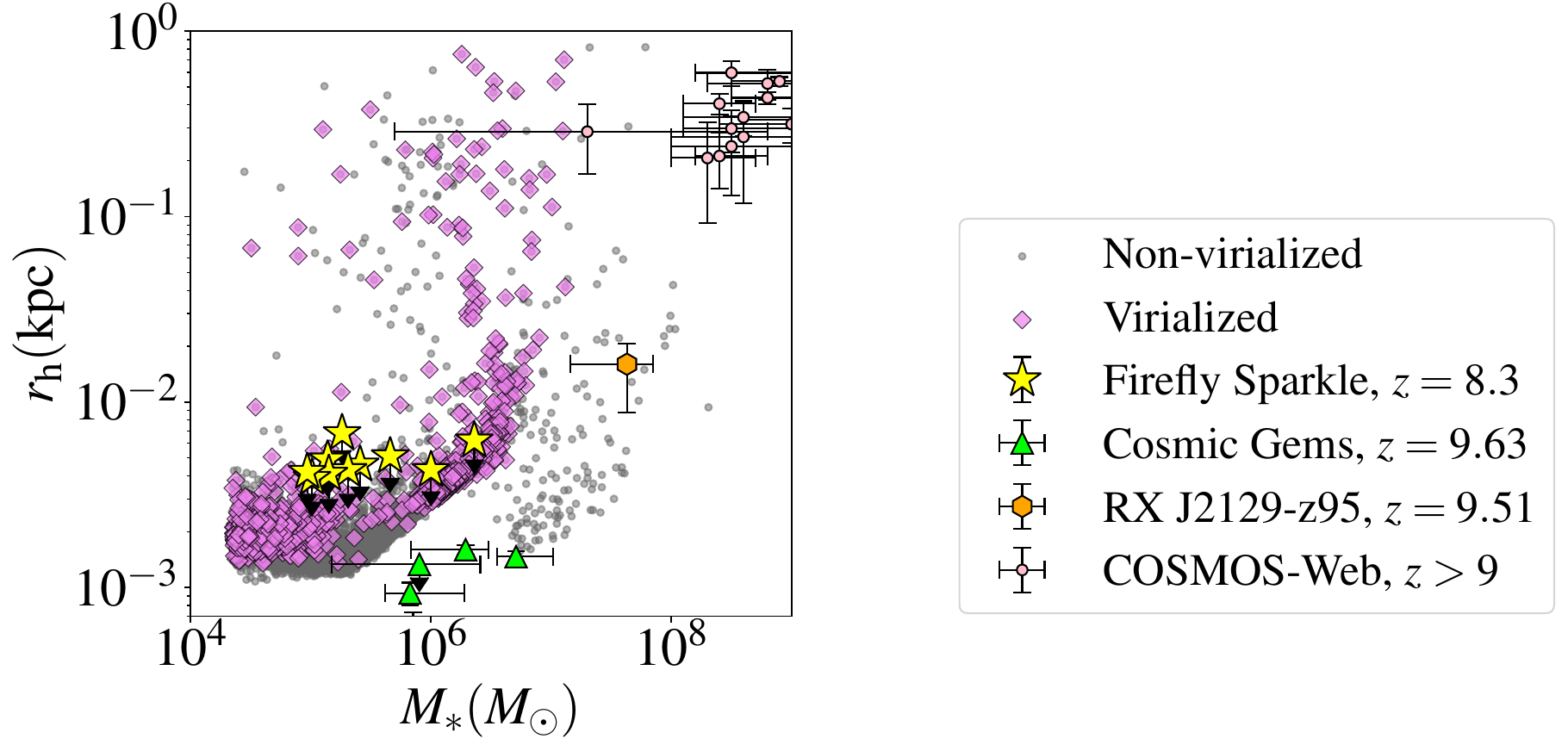}
    \includegraphics[width=0.42\textwidth]{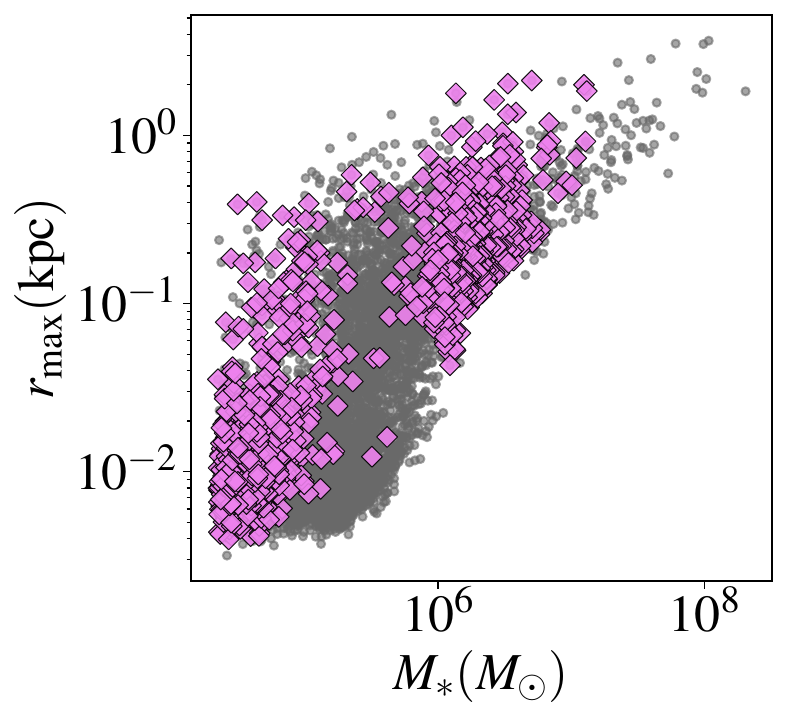}
    \includegraphics[width=0.42\textwidth]{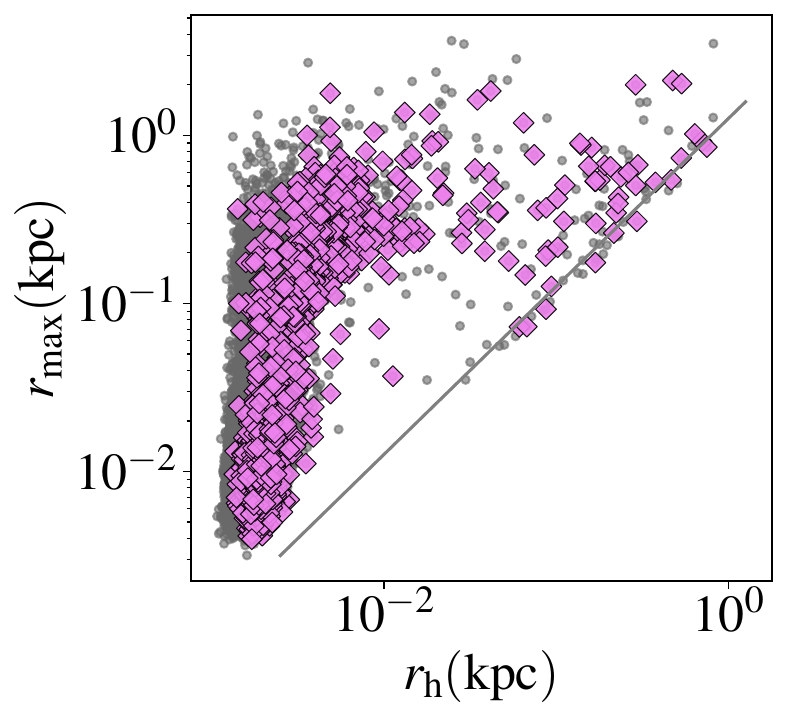}
    \caption{Top panel: Half mass radius vs. stellar mass. Grey points are non-virialized objects, and violet diamonds are virialized objects. 
    Bottom left panel: Maximum radius versus stellar mass. 
    Bottom right panel: Maximum radius vs. half mass radius. 
    The grey line shows the expected relation for a sphere of uniform density.
     All points shown are for stars+gas primary FOF run, at $z=12$. 
     Compact systems from high-redshift observations are overplotted on the top panel according to the legend in the top right (\citealt[][]{Mowla+24}, \citealt[][and private communication]{Adamo+24}, \citealt{franco_unveiling_2024,williams_magnified_2023}). } 
    \label{fig:radiim}
\end{figure*}
As mentioned in \S~\ref{sec:methods}, an accurate accounting of the density of star clusters requires a search for baryonic particles. 
With the structures in hand, we continue with an estimation of their compactness. 
For the purposes of the analytic arguments, we assumed a constant density, using $r_{\rm max}$ to estimate density (see Fig.~\ref{fig:radiim}, bottom left panel). 
However, to estimate the compactness and cuspiness of these stellar systems, we also compute the half-mass radius $r_{\rm h}$. 
This is the distance of the particle which encloses half the mass of the star cluster. 
Comparing $r_{\rm max}$ to $r_{\rm h }$ in Fig.~\ref{fig:radiim}, we see that the objects are certainly not of uniform density.
This contributes to our simulated timescales scattering outside of the errorbars in Fig.~\ref{fig:data-tscales}.
Thus, to calculate the surface density of the inner star cluster, which is most likely to be found by high redshift observations given the steep relation between surface brightness and redshift, we use $r_{\rm h}$ (see Fig.~\ref{fig:radiim} right panel). 

\begin{figure*}%
    \centering
    \includegraphics[width = 0.75\textwidth]{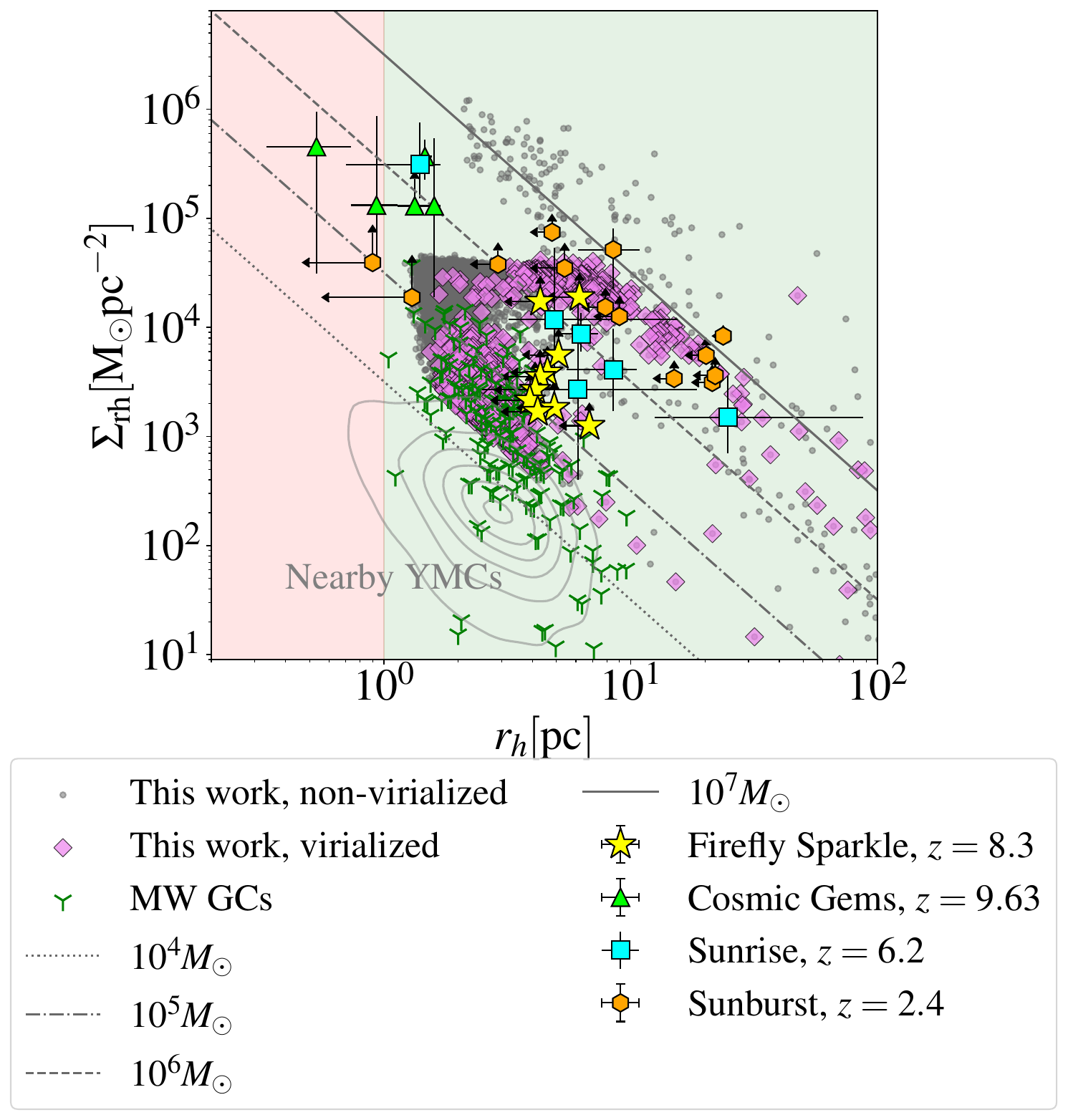}

    \caption{Surface density within half-mass radius versus half mass radius of simulated objects, colored by the logarithm of the baryonic density. Grey lines show the uniform density relations for various masses. The points shown are star+gas primary objects at $z=12$. Additionally, JWST high redshift clusters are shown for the Firefly Sparkle \citep[][]{Mowla+24}, Cosmic Gems \citep[][and private communication]{Adamo+24}, Sunrise \citep[][]{Vanzella+23}, and Sunburst \citep[][]{Vanzella+22} lensed arcs. Local MW globular clusters \citep[GCs;][]{Gieles+11} are shown in green, while the grey contours show local young massive clusters in nearby galaxies \citep[YMCs;][]{Brown+14}. 
    The green region on the right shows systems that are resolved in the simulation, whereas the red shaded region signifies the parameters space that cannot be explored in our simulation due to resolution.}%
    \label{fig:surface density}%
\end{figure*}

In Fig.~\ref{fig:surface density}, we plot the surface density of star clusters in the Stars$+$Gas primary run against their half-mass radius. 
Grey lines show the constant density relation for a given mass. 
We overplot recent {\it JWST} observations in the large neon points--clusters from the Firefly Sparkle, Cosmic Gems, Sunrise, and Sunburst lensed arcs \citep{Mowla+24, Adamo+24, Vanzella+22, Vanzella+23}. 
(We are limited by resolution on the left side of the plot. This is denoted by the red region, which shows radii smaller than the scale of gas cells in our simulation, whereas the green region is well-resolved.) 
Our points generally lie in the same region of parameter space as these unusual, high surface density systems.  
This is emphasized by the grey contours, showing local Young Massive Clusters \citep[YMC;][]{Brown+14} and local Globular Clusters \cite[GCs;][]{Gieles+11}, which fall in the lowest surface density regions of the plot. 
While some of our low surface density points lie in this region, we expect that inclusion of feedback processes would be required to reproduce the densities of these populations. 
The agreement between our simulated points and the high-z systems leads us to suggest that these observed systems formed in a very weak feedback scenario. 
We note that had we only investigated our simulations from a DM halo perspective, taking the surface density across the entire halo, our simulations would not agree with these points. 

We additionally compare our simulated star clusters at $z=12$ to a sample of star clusters observed with JWST (\citealt{claeyssens_tracing_2025,messa_anatomy_2025,Mowla+24,Adamo+24,fujimoto_primordial_2024,claeyssens_star_2023,Vanzella+23,vanzella_early_2022,Mayer+24}). 
The observed cluster surface density increases with redshift.
The median of our simulated dataset aligns slightly above the median observed by {\it JWST} at $z>8$.
Our systems have a lower density tail that reaches a much lower surface density than the whiskers of the observed distribution. 
However, we note that these low density systems would almost certainly not be observed at high redshift. 
Furthermore, those low density systems may be the most susceptible to feedback processes, and may not form in a full-feedback simulation. 
\begin{figure}
    \centering
    \includegraphics[width=0.95\linewidth]{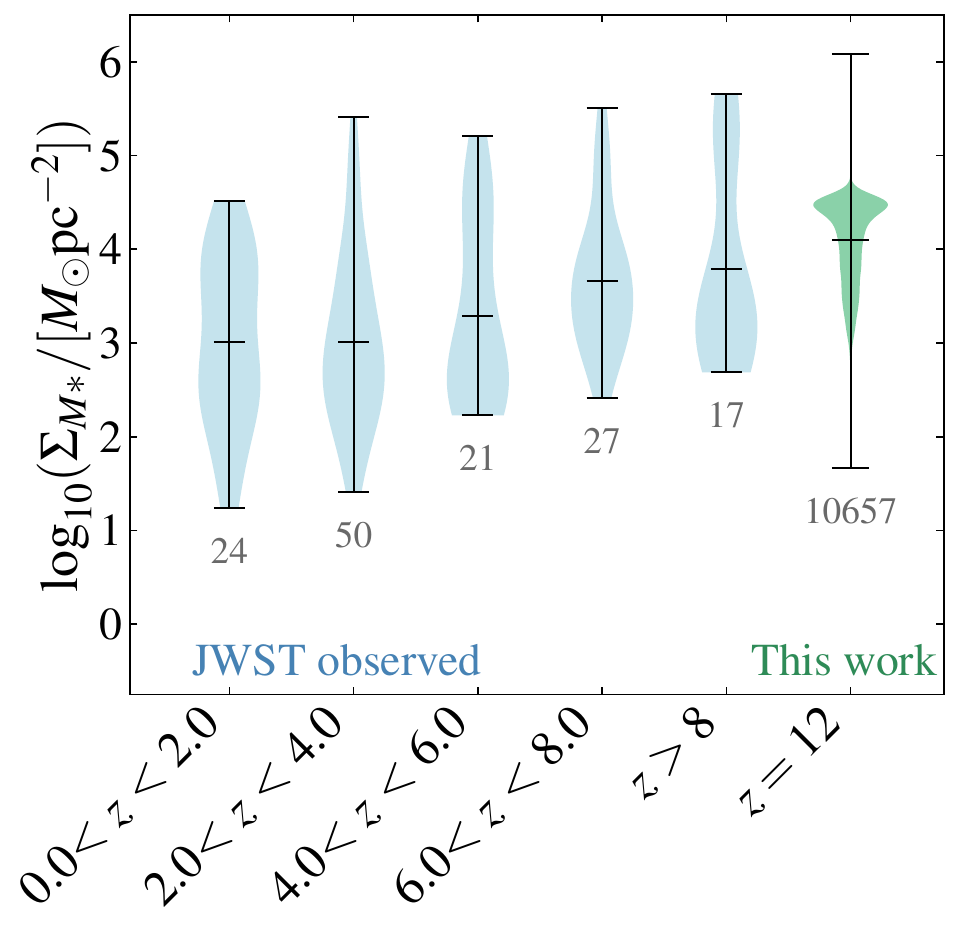}
    \caption{Violin plots of surface density of JWST observed star clusters (blue, left side) compared to our simulated sample (green, right side). The mean, minimum, and maximum of each distribution are shown in the whiskers. For observed systems, clusters with effective radius $<25$pc were selected. The simulated structures were filtered for those with $M_*>10^{4.368}M_\odot$, the minimum mass of the observed sample, and radius $<25$ pc, to match the cutoff in assembling the observed sample.  Below each distribution the grey number is the number of objects included in the distribution. 
    The sample includes clusters from \cite{claeyssens_tracing_2025,messa_anatomy_2025,Mowla+24}; \cite{Adamo+24} and private communication; \cite{fujimoto_primordial_2024,claeyssens_star_2023,Vanzella+23,vanzella_early_2022} }
    \label{fig:redshift-sdensity}
\end{figure}

\section{Discussion and Conclusions}
\label{sec:summary}

In this work, we present star clusters in a cosmological simulation, grouped using  structure-finding methods focused on multiple particle types. 
We compare the properties of clusters grouped by the typical Dark Matter-centric halo finding algorithm with a version focused on gas and star particles aimed at mimicking an observational approach.

In general,  
the star clusters
produced by our simulation
are of comparable size and surface density to observed systems at high redshift, although we do not have the resolution to accurately model clusters below the pc scale.
An important caveat is that although our simulation does not include stellar feedback, feedback processes should be important in shaping the distribution of clusters that form.
For this reason,
we propose that {\it these simulations are an upper limit for the most extreme star formation permitted for $\Lambda$CDM Pop III stars}.
Our simulated systems may have similar properties to observed systems which form in high star formation efficiency cases.
Using the methods available here, it is difficult to separate the role of feedback from other physical processes in shaping the high stellar densities in this era. High densities may be attributed to weak feedback, efficient cooling, small sizes of baryonic objects, and possibly other factors.
We speculate that the inclusion of feedback may have several effects. 
Firstly, low-density systems may be disrupted, removing the low-density tail of systems (seen in the violin plot of Fig.~\ref{fig:redshift-sdensity}).
Furthermore, the injection of kinetic energy and momentum provided by feedback could increase the virial radius of clusters.
Additionally, it is possible that the most massive and high-density clusters (non-virialized systems that lie around the $10^7\, M_\odot$ line in Fig.~\ref{fig:surface density}) may be prevented from reaching such high masses. 
This could potentially move these systems towards slightly lower mass in the Figure, perhaps driving them to emulate the Cosmic Gems systems. 
However, given recent studies suggesting that at $\sigma>10^3 M_\odot/\text{pc}^2$ the efficient conversion of gas into stars is inevitable \citep[e.g.,][]{menon_outflows_2023,grudic_when_2018}, we suggest that this remains uncertain.
Finally, if strong feedback processes were implemented, this could lead to a reduced occurrence of star clusters outside of halos, thus lessening the discrepancy between the DM halo-finding algorithm and the baryonic version.

Clusters at cosmic dawn serve as the building blocks of early galaxies and may later evolve into galactic nuclei or globular clusters.
While we do not explore the chemical properties of these clusters or their evolution to lower redshifts in this work, their potential role in shaping early galaxy populations remains significant. 
Lensed, high-redshift clusters are often inferred to have young ages \citep[e.g.,][]{Adamo+24}, particularly when compared to the time elapsed between our simulations at $z=12$.
These observed Pop II systems likely represent a later generation of stars, whereas our simulated clusters correspond to an earlier population present at $z\geq12$. 
If the weak-feedback, low-stream-velocity scenario explored here applies at high redshifts, the clusters we study may influence broader properties of the early galaxy population—even if individual clusters are not directly observed at $z=12$.
Through mergers, accretion, and continued growth, these systems become constituent components of JWST's highest redshift galaxies.
The extreme stellar densities of these early clusters could contribute to unusual chemical enrichment patterns \citep[e.g.,][]{charbonnel_n-enhancement_2023}.
Moreover, \cite{Menon+24} demonstrate that high-surface-density star-forming clumps create conditions favorable for Lyman continuum (LyC) photon escape during the lifetimes of massive stars, suggesting that these clusters may play a role in driving cosmic reionization.

Finally, these dense star clusters serve as ideal sites for early black hole formation in the intermediate mass regime. 
If the recoil kick experienced during a black hole merger is lower than the system escape velocity (which may often be the case given the low spin reported by LIGO/Virgo; e.g.,  \citealt[][]{collaboration_gwtc-3_2023,collaboration_population_2023,collaboration_binary_2019}), the remnant will remain in the cluster and has the chance to interact and merge again \citep[e.g.,][]{rodriguez_black_2019}. 
Through a gravitational runaway process, this may result in intermediate mass black holes (IMBHs) of $10^{3-5} M_\odot$ \citep[e.g.,][]{sakurai_growth_2019,sakurai_formation_2017,hirano_one_2014,zwart_runaway_2002}. 
Additionally, if these clusters merge into a larger protogalaxy, dynamical friction could facilitate the formation of early overmassive supermassive black holes (SMBHs) \citep[e.g.,][]{dekel_growth_2024,Mayer+24}. 
This population of IMBHs at high redshift would provide sites of dynamical interaction that could lead to stellar mass binary black hole mergers and extreme mass ratio inspirals \citep[e.g.,][]{xuan_detecting_2024,arca_sedda_quiescent_2023,naoz_combined_2022,aarseth_mergers_2012}, as well as IMBH mergers and wandering black holes through the interaction of star clusters with larger halos and galaxies \citep[see e.g.,][]{greene_2020_intermediate}. 
This would imply many promising avenues of detection through next generation gravitational wave detectors.

\paragraph{Summary} Our main results are as follows: 
\begin{enumerate}
    \item When tracking star clusters, rather than dark matter halos, a discrepancy is found in the number counts by stellar mass and star formation rate up to a factor of several at all masses. In these cases, studies should consider which method most benefits the observable quantity of interest. Often, the star primary or star and gas primary algorithm will identify concentrations of starlight, which are useful for observational comparison. 
    \item We study the virialization of star clusters in our simulation in detail, identifying the processes that lead systems at a given mass and initial density to equilibrate or not. 
    For high redshift clusters and galaxies, achieving virialization erases the system's initial conditions in the hierarchical process of galaxy buildup. 
    We show that many systems at the redshifts considered here are not virialized. 
    \item In particular, we find two populations of virialized clusters originating from the tracks of molecular cooling of primordial metal-free gas. They correspond to roughly $10^4-10^5M_\odot$ and $10^6-10^7 M_\odot$. We show how hierarchical mergers disrupt these objects, providing an analytical framework for virialization and disruption.  
    \item We use our star and gas primary cluster searches to explore the stellar surface density of high redshift clusters. 
    Our results are consistent with {\it JWST}'s highest surface density clusters. 
    Since the simulation suite investigated here does not include stellar feedback, this could imply that these {\it JWST} clusters formed in a low-feedback mode. 
    For example, our results would be consistent with ``feedback-free" star clusters \citep[e.g.,][]{dekel_efficient_2023}. 
    In any case, the clusters shown here represent a theoretical upper limit for extreme star formation permitted within $\Lambda$CDM Pop III stellar systems. 
    
\end{enumerate}

This work highlights the importance of detailed investigation of $\Lambda$CDM simulations to uncover the relationship between simulated structures and observed systems and hints that a weak feedback mode may be present in the early Universe. 

\section*{Acknowledgements}

 C.E.W.  acknowledges the support of the National Science Foundation Graduate Research Fellowship, the University of California, Los Angeles (UCLA), and the UCLA Center for Diverse Leadership in Science Fellowship. C.E.W., W.L., S.N., Y.S.C, B.B., F.M., and M.V. thank the support of NASA grant Nos. 80NSSC20K0500 (9-
ATP19-0020) and 80NSSC24K0773 (ATP-23-
ATP23-0149) and the XSEDE/ACCESS AST180056 allocation, as well as the UCLA cluster Hoffman2 for computational resources. B.B. also thanks the the Alfred P. Sloan Foundation and the Packard Foundation for support.
N.Y. acknowledges financial support from JSPS International Leading Research 23K20035.
 F.M. acknowledges support by the
European Union—NextGeneration EU within PRIN 2022
project n.20229YBSAN—Globular clusters in cosmological
simulations and in lensed fields: from their birth to the present
epoch.
 Simulation runs for this work were performed on the Anvil Cluster \citep{song_anvil_2022}.
This material is based upon work supported by the National Science Foundation Graduate Research Fellowship Program under Grant No. DGE-2034835. Any opinions, findings, conclusions, or recommendations expressed in this material are those of the author(s) and do not necessarily reflect the views of the National Science Foundation.
This work used computational and storage services associated with the Hoffman2 Cluster which is operated by the UCLA Office of Advanced Research Computing’s Research Technology Group.

{\bf Data Availability:}
A catalog of the star clusters and dark matter halos used in this work is available for public use: \href{https://www.astro.ucla.edu/~clairewilliams/cluster-catalog}{Link}.
The code to post process the FOF runs is also available on GitHub: \href{https://github.com/astro-claire/process-fof.git}{Link}.

%


\software{astropy \citep{2013A&A...558A..33A,2018AJ....156..123A}, yt \citep[][]{turk_yt_2011}, matplotlib \citep{Matplotlib},  numpy \citep{harris2020numpy}, scipy \citep{2020SciPy-NMeth}.
          } 



\appendix

\section{Appendix: Convergence Tests}
\label{app:convergence}
In order to ensure convergence  and assess the robustness of our structure finding algorithm we test for convergence by varying the linking length used in the FOF algorithm. 
During the particle linking stage, two particles are identified as group members if they are closer to one another than $\ell_{\rm FOF}$ times the mean inter-particle separation. 
By default, we set $\ell_{\rm FOF}=0.2$. 
\begin{figure}
    \centering
    \includegraphics[width=0.5\linewidth]{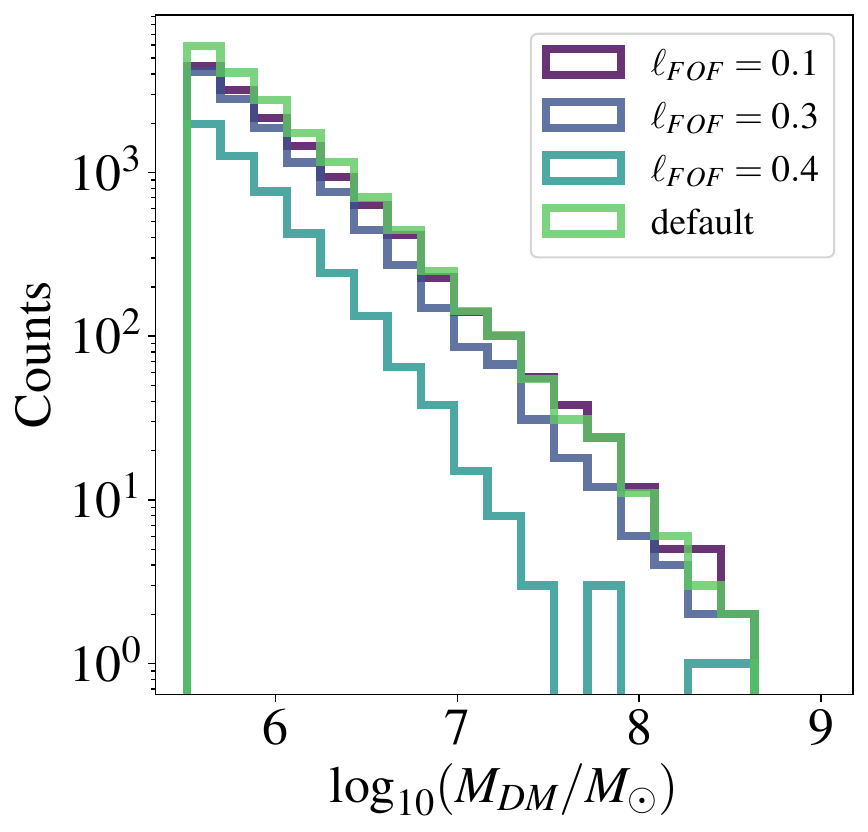}
    \caption{Counts of objects by DM mass in DM primary FOF run with varying linking lengths ($\ell_{\rm FOF}$). }
    \label{fig:convergence}
\end{figure}

We re-ran our DM-primary, stars$+$gas secondary structure finding algorithm with three additional values of the linking length: $\ell_{\rm FOF} = 0.1, 0.3,$ and $0.4$. 
The counts by dark matter halo mass are plotted in Fig.~\ref{fig:convergence}. 
The simulated number of objects is consistent for variations within $0.1$ of our value, and starts to diverge when the linking length is varied by a factor of $2$. 
The variations between these convergence runs are much smaller than the error found between the DM-centric and baryon-centric methods, so we can confirm that our simulated discrepancies are not due to convergence issues.  

\section{Appendix: Cluster Mass Function}
\label{app:clustermassfunction}

In this Section we provide the cumulative cluster mass function for structures in our simulation box. 
Since we have increased $\sigma_8$ in our simulation (see discussion in \citealt{Lake+23b}, Section~2), this mass function is only applicable to high density peaks in the Universe.  Therefore, our structures are more numerous than the average patch of the Universe with similar volume.

\begin{figure}
    \centering
    \includegraphics[width=0.5\linewidth]{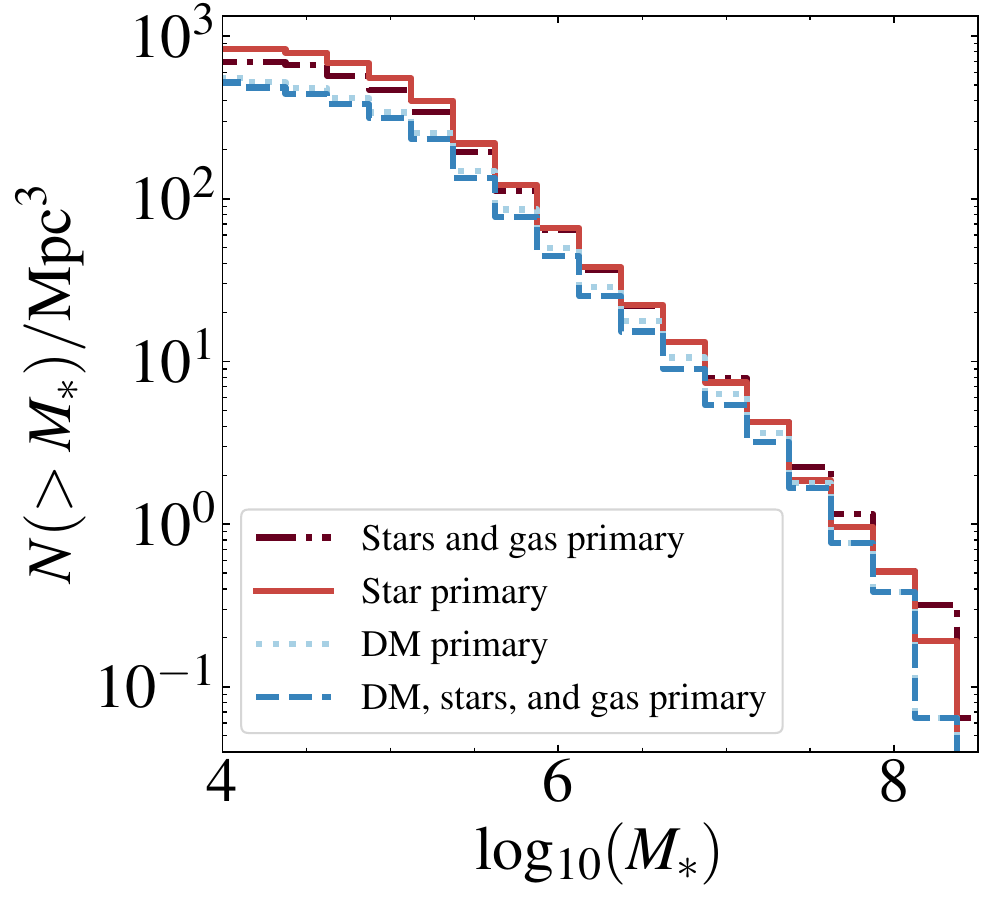}
    \caption{Cumulative mass function of star clusters in the simulation. The volume has not been corrected for increased $\sigma_8$, so the provided data represents a very high density peak in the Universe. Results are presented at $z=12$. The line styles follow the convention defined in Fig.~\ref{fig:mstarhistz12}.}
    \label{fig:mass_function}
\end{figure}

\section{Appendix: Ancillary data for dynamical investigation }

\subsection{Analytic Timescales}
\label{app:analytictimescales}
\begin{figure*}
    \centering
    \includegraphics[width=0.48\textwidth]{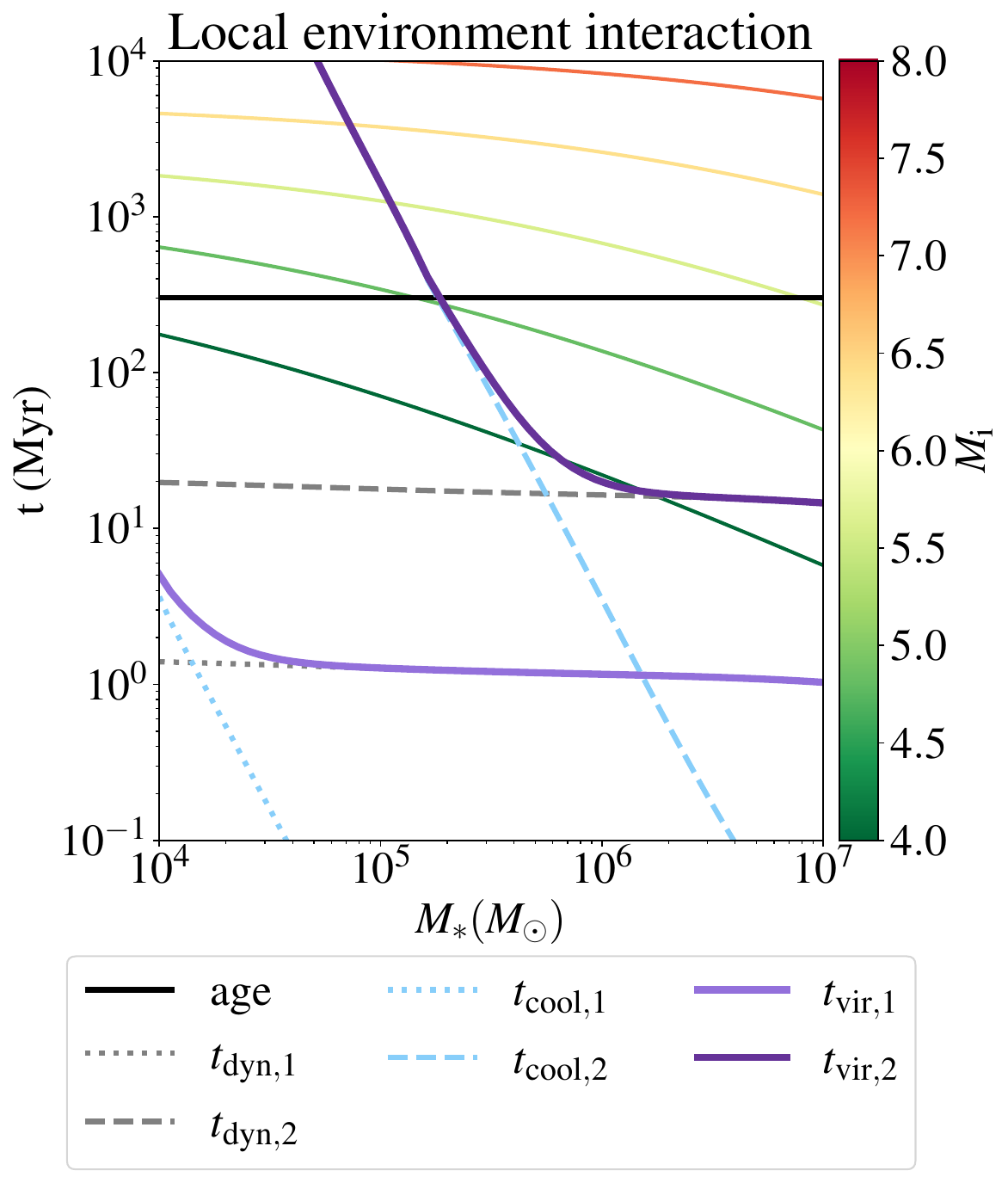}
\includegraphics[width=0.48\textwidth]{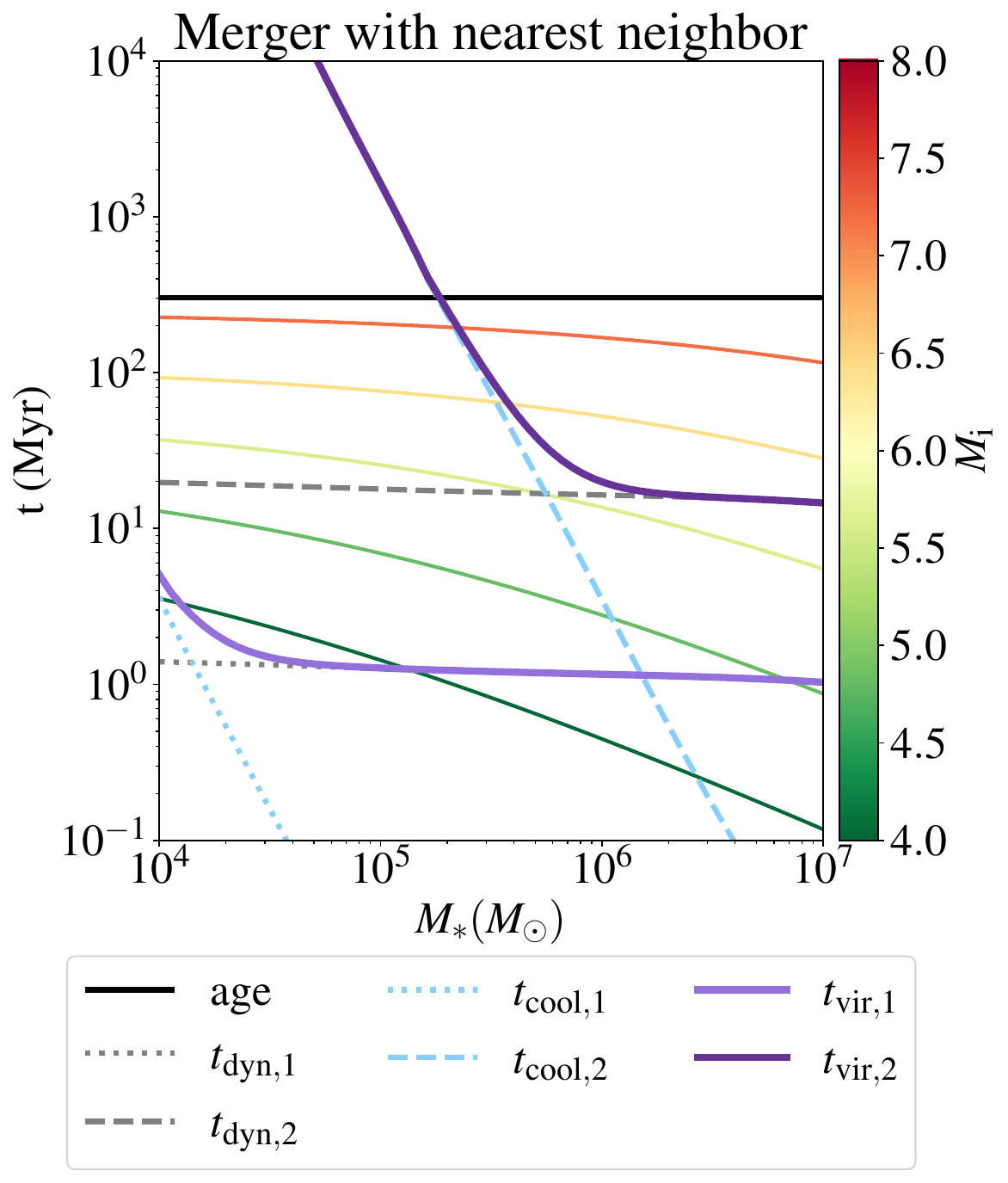}
\caption{Comparison of analytic timescales for virialization (purple lines) and interaction (tan line). The blue lines show the H$_2$ cooling timescale for the two typical densities discussed in \S~\ref{sec:gasconditions}. The grey lines show the dynamical timescales for those densities. The solid black line is the age of the universe at $z=14$. On the LHS, the color bar separates lines of interaciton timescale by interacting halo mass, with the timescale calculated for the average nearby environment (see Eq.~\ref{eq:tinteract} and discussion in text). On the RHS, the interaction timescales are instead calculated as if the interacting halo mass was at the distance of the nearest neighbor in the simulation, providing a lower limit. }
    \label{fig:analyticaltscales}
\end{figure*}

In Fig.~\ref{fig:analyticaltscales}, we show a more detailed comparison between the timescale for virialization and the timescales for interaction. 
This figure shows how the upper and lower limit error bars in Fig.~\ref{fig:data-tscales} were derived. 
In both panels, the virialization timescale is shown in purple as in  Fig.~\ref{fig:data-tscales}, along with the cooling and dynamical time for each density (see Eq.~\ref{eq:virialization}). 
The black horizontal line shows the age of the Universe. 
The color bar denotes the mass of the interacting halo (used in the number density calculation, Eq.~\ref{eq:numberdens}). 

On the left, we show upper limit case for interaction--this is if we use the average density of halos in the local region to calibrate the number density. 
This is almost certainly an overestimate for most systems, since here we assume uniformly distributed halos while real systems are highly clustered. 
Thus, to provide a lower limit for the interaction timescale, we show the right hand panel with the timescale calibrated to the interaction time of the nearest neighboring halo.
This overestimates the interaction timescale in most cases, because there is not a constant supply of neighbors at the distance of the closest neighbor, and many system's nearest neighbor is much farther than the simulation box average (see Fig.~\ref{fig:numberDM}). 

\subsection{Numerically-derived quantities: data and fitting functions}
\label{app:simulatedquantities}
Here we provide the simulated data used to calibrate the analytic timescales from our investigation.

\begin{figure*}
    \centering
    \includegraphics[width=\textwidth]{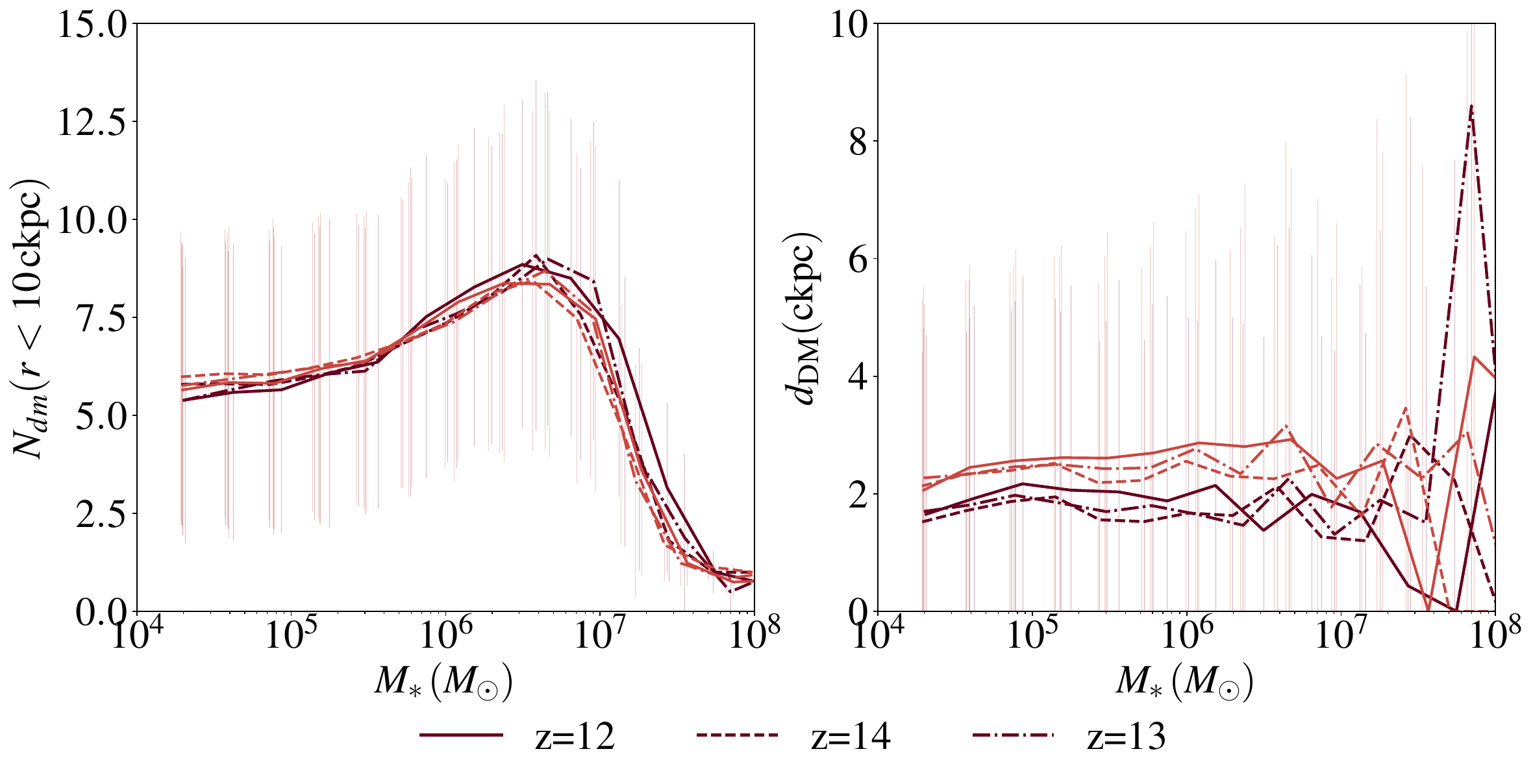}
    \caption{Left panel: Number of DM neighbors within a radius of 10 comoving kpc as a function of stellar mass for various redshifts. The linestyles denote the various redshifts of the simulation snapshots, while the lighter red is star primary and the black is star + gas primary (all redshifts and FOF runs are similar). The error bars show one standard deviation, $\sigma$, above and below the mean in each bin. Right panel: Mean distance to the nearest dark matter halo.}
    \label{fig:numberDM}
\end{figure*}

Figure~\ref{fig:numberDM} shows simulated data relating to the dark matter environment of baryon-centric structures.  
In the left panel, the average number of dark matter halos from the DM-centric catalog located within 10 ckpc of each baryon-centric object is plotted as a function of stellar mass. 
On the right, the distance to the nearest DM halo is plotted again as a function of stellar mass for the baryon-primary objects. 
In both panels, the error bars denote one standard deviation above and below the mean in each bin. 

\begin{figure}
    \centering
    \includegraphics[width=0.48\textwidth]{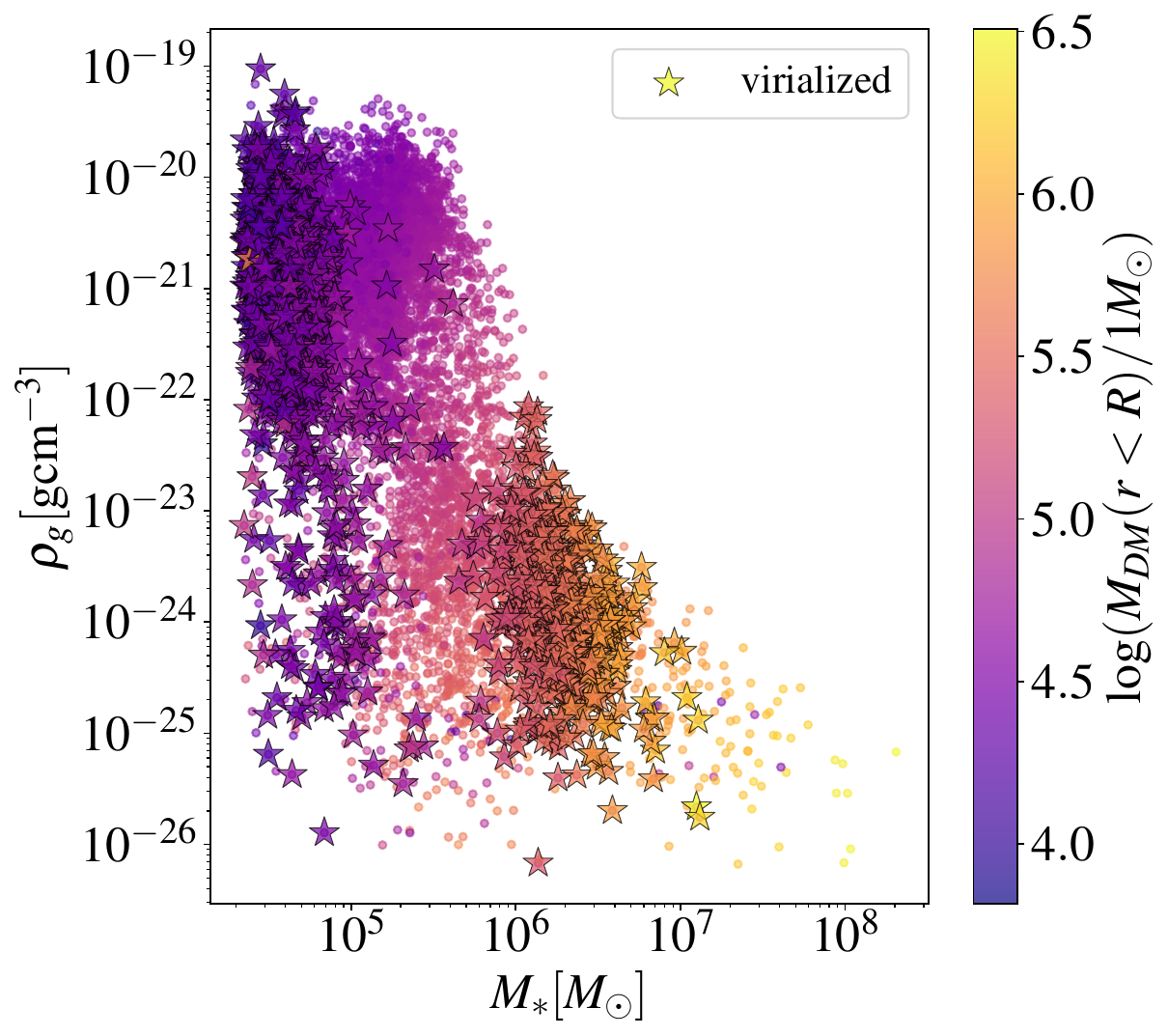}
    \caption{Mean gas density vs. stellar mass for simulated objects in the stars+gas primary run at $z=12$. The color bar shows the dark matter mass within the radius of the object.  }
    \label{fig:rho_m}
\end{figure}

Figure~\ref{fig:rho_m} gives the average gas density as a function of stellar mass for objects detected in the stars$+$gas primary catalog.
The two populations of virialized objects are highlighted. 
This average gas density is computed using the mass contained in gas cells within a sphere of radius equal to the maximum galactocentric star particle distance. 
Reassuringly, the densities estimated for the analytic argument from the high redshift DM halos (see Table~\ref{tab:derivedvalues}) fall roughly within the two typical gas densities in the simulation at this later time.

\begin{figure*}
    \centering
    \includegraphics[width=\textwidth]{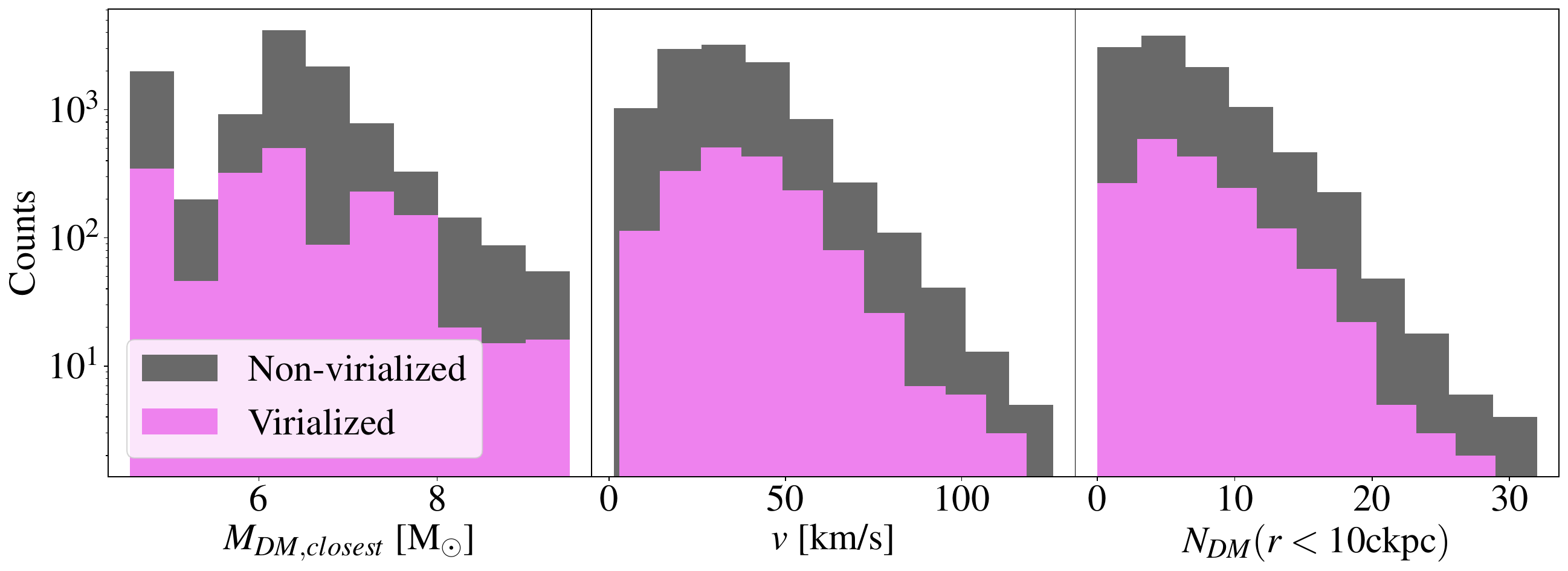}
    \caption{Left panel: Mass of nearest neighbor in solar masses. Central panel: group velocity with respect to the bulk flow. Right panel: Number of DM halos within 10 ckpc, as traced by the DMP FOF run. Grey histograms represent the non-virialized objects, and violet represent the virialized objects. All results shown are for the stars+gas primary run at $z=12$. Mean values are listed in Tab~\ref{tab:derivedvalues}. }
    \label{fig:velocityhist}
\end{figure*}

Next, we show additional histograms of quantities used for the interaction timescale in Fig.~\ref{fig:velocityhist}. 
From left to right, the panels show the mass of the closest dark matter object, the group velocity with respect to the bulk flow, and the number of dark matter neighbors within 10 ckpc. 
The populations are split into non-virialized and virialized. 
In general, the virialized and non-virialized populations trace each other. 
However, there are a dearth of virialized objects with a large number of DM neighbors, and a smaller fraction of virialized objects with the most massive DM neighbors. 
This aligns with the analytical expectation--a large number of neighbors or an extremely massive neighbor is more likely to disrupt a system's state of equilibrium. 

\paragraph{$M_*-M_{\rm DM}$ relation}

We find a good fit to the simulated data with $R^2>0.91$ up to $M_* \leq 10^7 M_\odot$ using the following fit: 
\begin{multline}
    \log M_{\rm DM} = -0.01584 (\log M_* )^4 +0.308989 ( \log M_*) ^3 -2.03908( \log M_*)^2+ 5.61641 \log M_* \\ +0.19979, 
\end{multline}
where $\log M_* = \log10 \left(\frac{M_{\rm *}}{1 M_\odot}\right)$ and  $\log M_{\rm DM} = \log10 \left(\frac{M_{\rm DM}}{1 M_\odot}\right)$. 
 
\paragraph{$M_* - M_{\rm g}$ relation}
Additionally, we use the following fit to model the stellar mass to gas mass relation: 
\begin{multline}
    \log M_{\rm g}  = -0.020562 (\log M_* )^4 +0.402479 ( \log M_*) ^3 -2.717265( \log M_*)^2+ 7.688866 \log M_* \\ -2.688908, 
\end{multline}
where $\log M_{\rm g} = \log10 \left(\frac{M_{\rm g}}{1 M_\odot}\right)$. 
This provides $R^2>0.83$ for all masses below $M_*\leq 10^7$
These fits are used to estimate the halo and gas mass as a function of stellar mass for the purpose of deriving the virialization timescale.

\section{Appendix: Results at various redshifts}
\label{app:redshifts}
\begin{figure*}
    \centering
    \includegraphics[width = 0.45\linewidth]{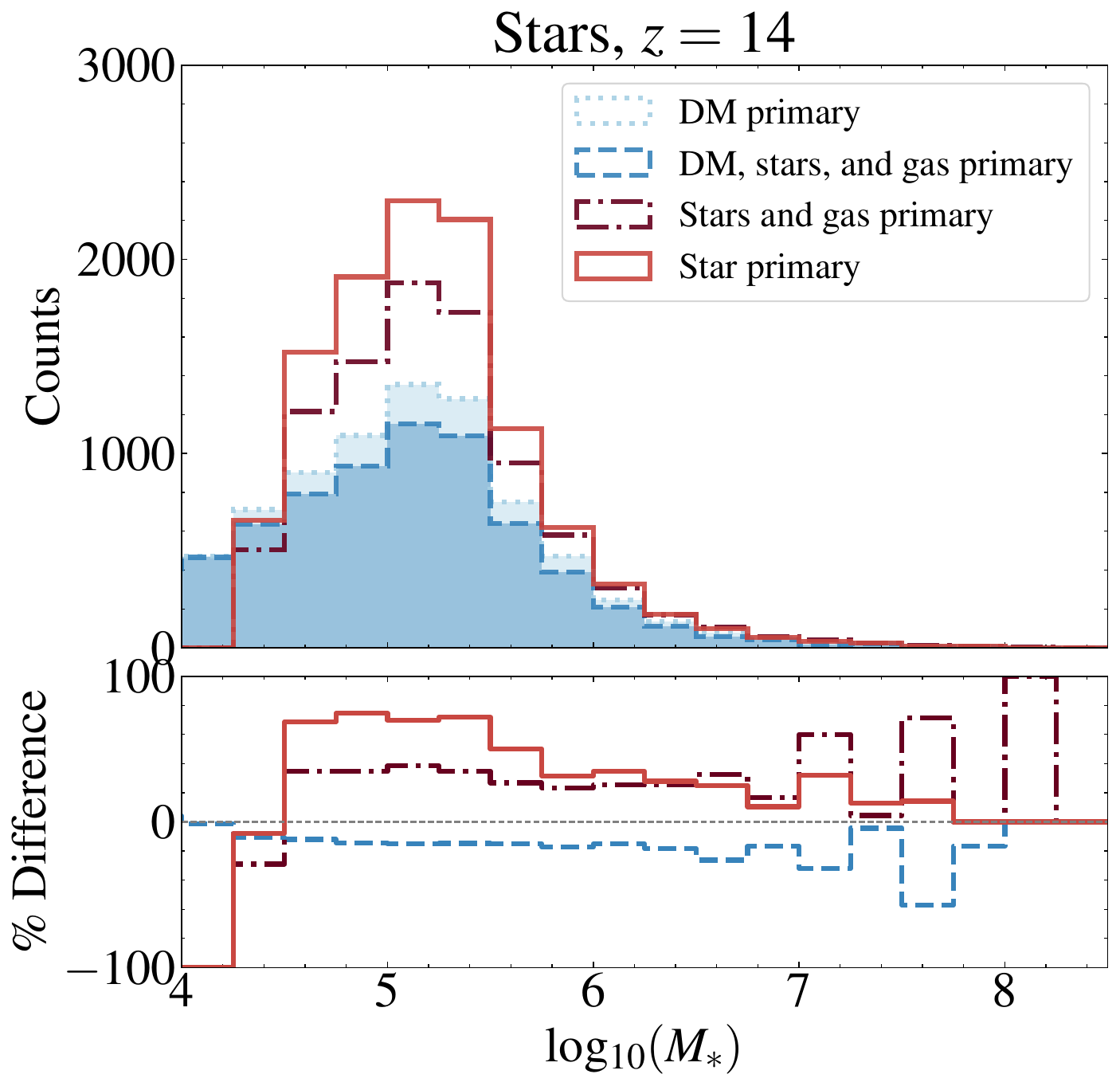}\includegraphics[width = 0.45\linewidth]{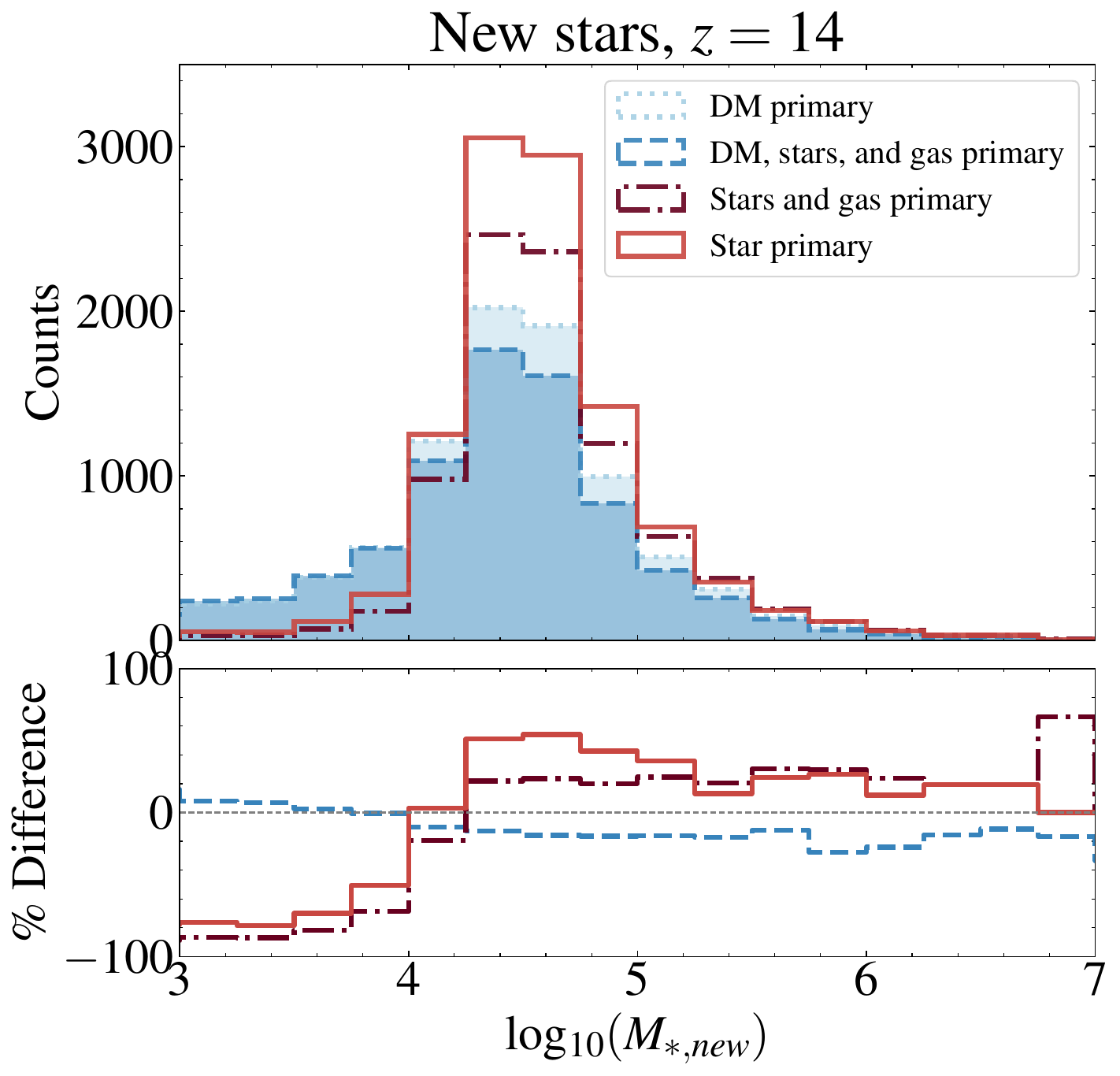}    
    \caption{Top left panel: Histogram of object counts by stellar mass at $z=14$ using algorithms centered on the particles listed in Table~\ref{tab:FOFruns}. Bottom left panel: Percent difference between each FOF run and the standard DM-primary FOF ($f= N/N_{\rm DMP}$). Grey horizontal line shows $f=1$. 
    Blue shades are runs that include DM--the DM primary (dotted) and the stars, gas, and DM primary (dashed). The red shades are baryonic runs--the star primary (solid) and the stars and gas primary (dot-dashed). Top right panel: Histogram of object counts by newly-formed stellar mass at $z=12$ using FOF primaries listed in Table~\ref{tab:FOFruns}. Bottom right panel: Percent difference between each FOF run and the standard DM-primary FOF ($f= N/N_{\rm DMP}$). Grey horizontal line shows $f=1$. 
    }
    \label{fig:mstarhistz14}
\end{figure*}

\begin{figure*}
    \centering
    \includegraphics[width = 0.45\linewidth]{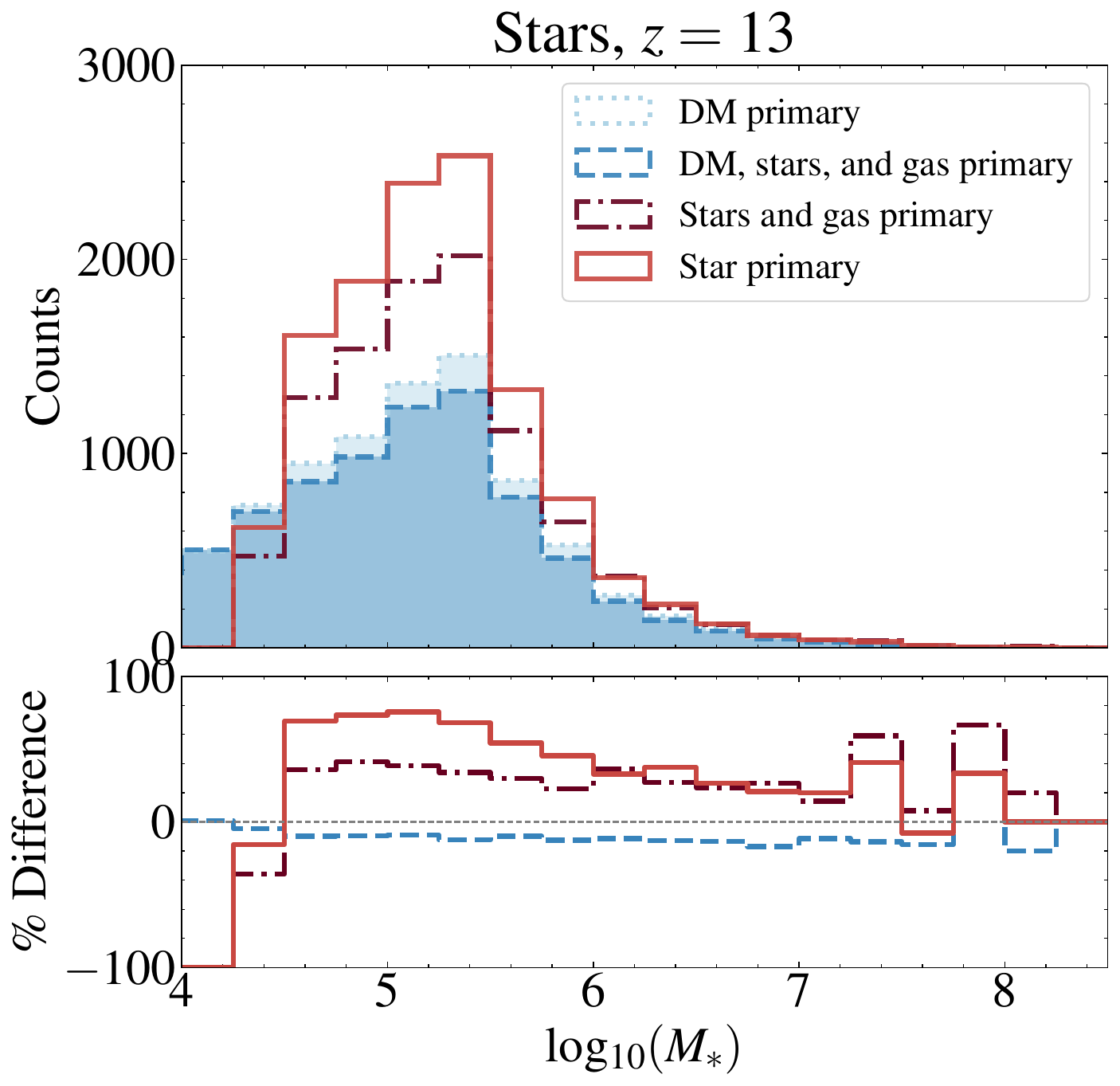}\includegraphics[width = 0.45\linewidth]{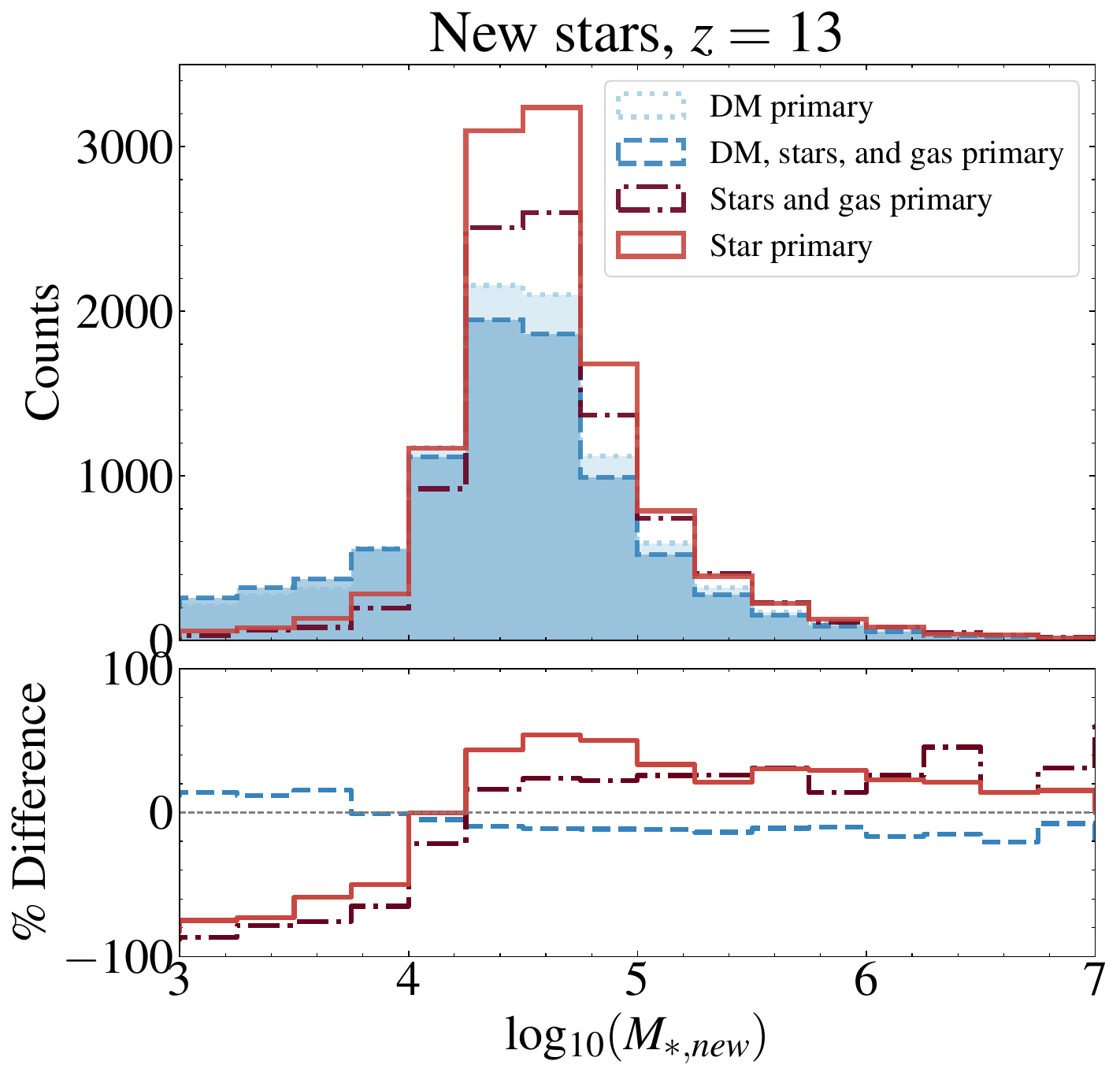}    
    \caption{Top left panel: Histogram of object counts by stellar mass at $z=13$ using algorithms centered on the particles listed in Table~\ref{tab:FOFruns}. Bottom left panel: Percent difference between each FOF run and the standard DM-primary FOF ($f= N/N_{\rm DMP}$). Grey horizontal line shows $f=1$. 
    Blue shades are runs that include DM--the DM primary (dotted) and the stars, gas, and DM primary (dashed). The red shades are baryonic runs--the star primary (solid) and the stars and gas primary (dot-dashed). Top right panel: Histogram of object counts by newly-formed stellar mass at $z=12$ using FOF primaries listed in Table~\ref{tab:FOFruns}. Bottom right panel: Percent difference between each FOF run and the standard DM-primary FOF ($f= N/N_{\rm DMP}$). Grey horizontal line shows $f=1$. 
    }
    \label{fig:mstarhistz13}
\end{figure*}

\begin{figure}%
    \centering
    \includegraphics[width = 0.32\linewidth]{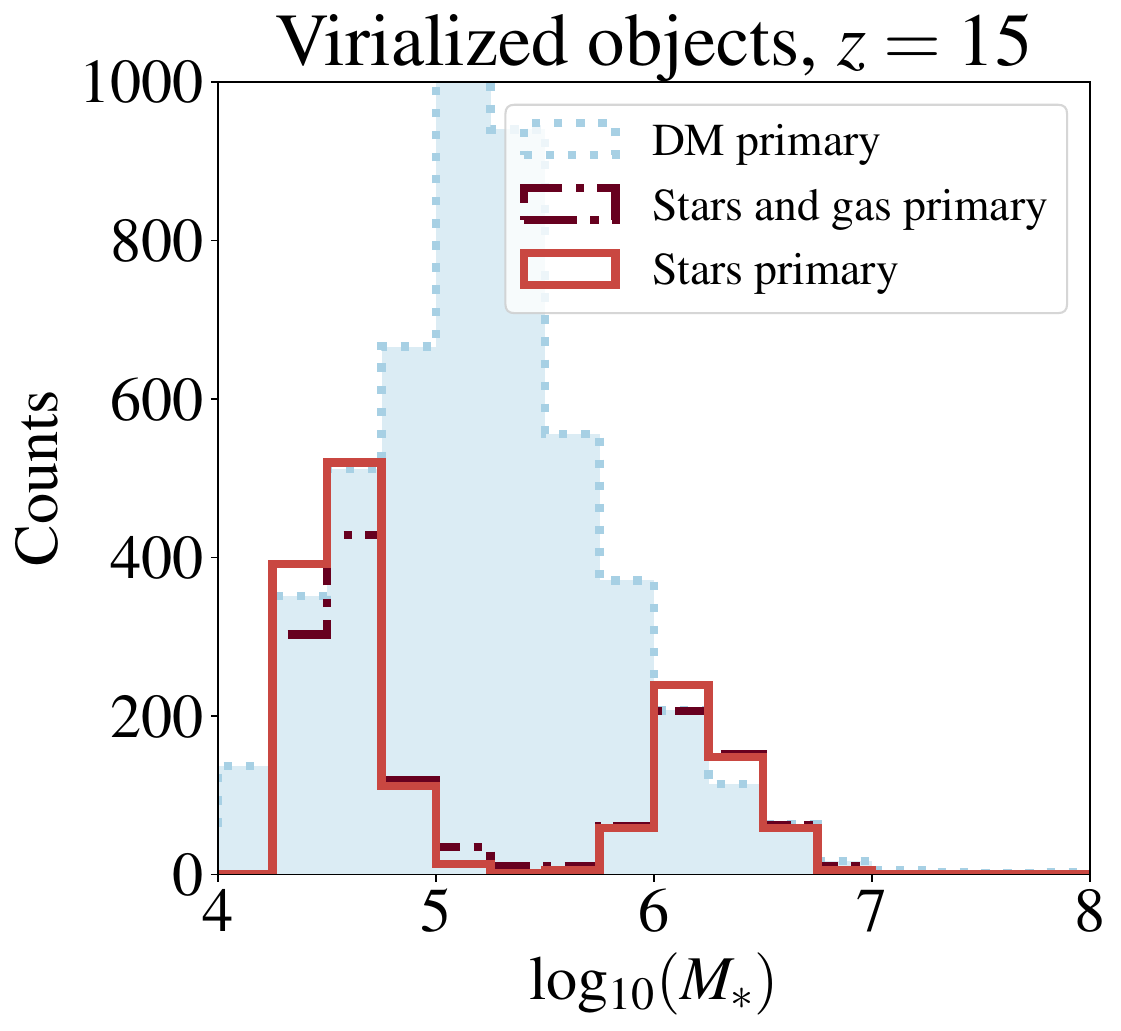}
    \includegraphics[width = 0.32\linewidth]{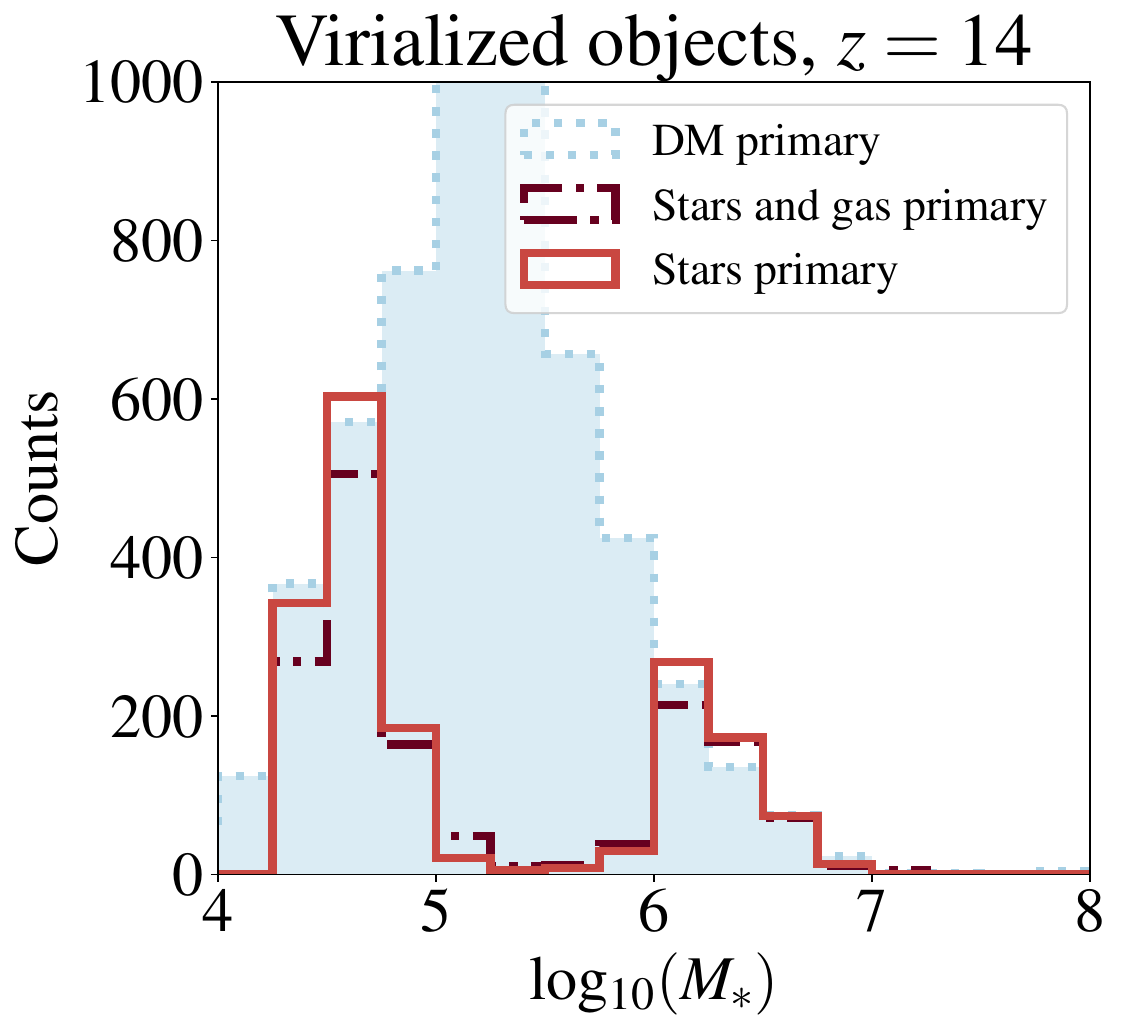}
    \includegraphics[width = 0.32\linewidth]{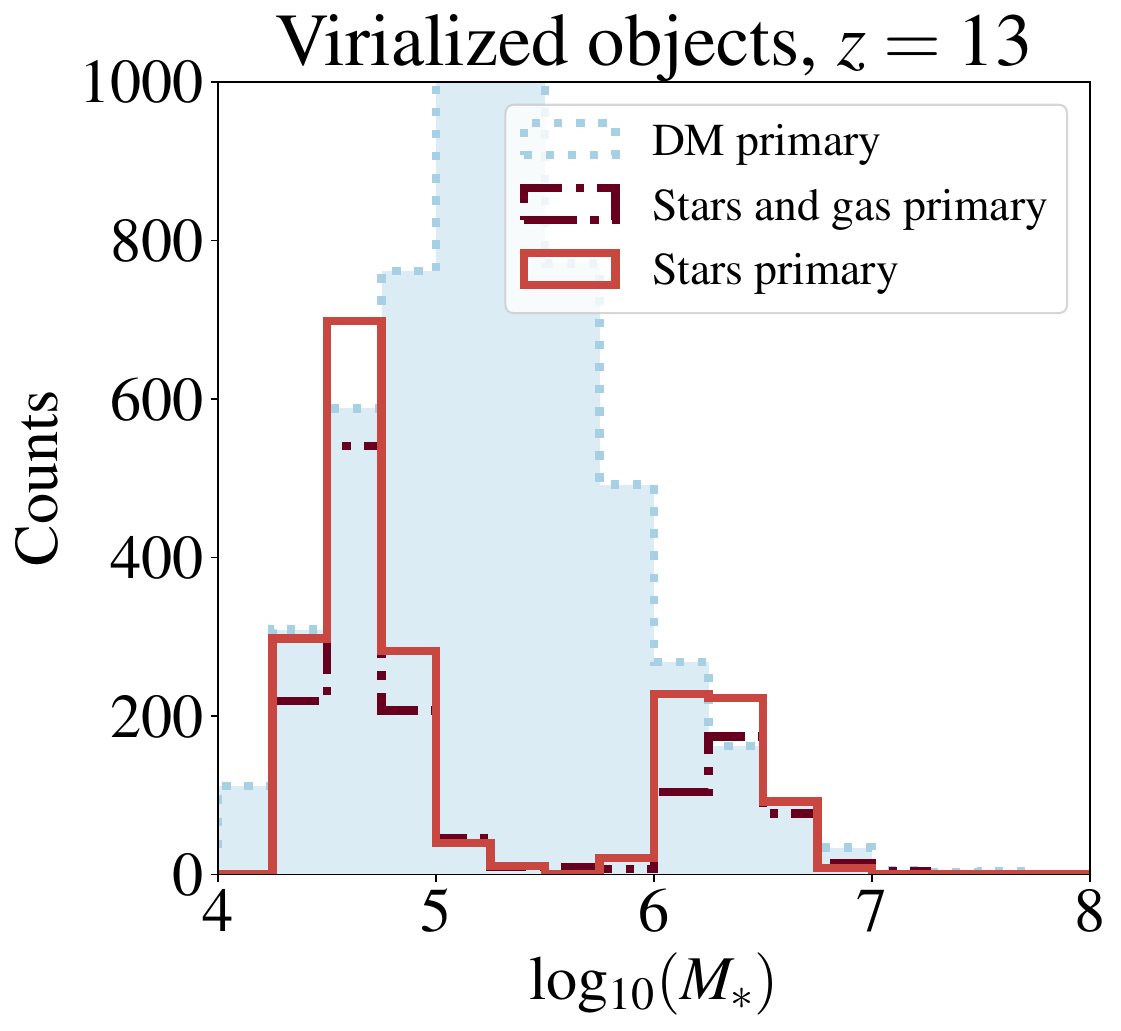}
    \caption{ Number of virialized objects at $z=15,14,$ and $13$ (from left to right) by stellar mass in solar masses. Light blue is the DM primary (dotted). The red shades are baryonic runs--the star primary (solid) and the stars and gas primary (dot-dashed).
    }%
    \label{fig:virhistredshifts}%
\end{figure}

In this section, we provide results at $13\leq z \leq 15$. 
Figures~\ref{fig:mstarhistz14} and~\ref{fig:mstarhistz13} give the overall histograms of objects by stellar mass as in Fig.~\ref{fig:mstarhistz12}. 
Figure~\ref{fig:virhistredshifts} shows the virialized population by stellar mass at $z = 13, 14,$ \& $15$, as in  Fig.~\ref{fig:fvir}. 
The trends discussed in \S~\ref{sec:boundclusters} are consistent at all the redshifts investigated here. 

\bibliography{main}{}
\bibliographystyle{aasjournal}



\end{document}